\documentclass[authoryear,preprint, 12pt]{elsarticle}
\usepackage{fullpage}
\usepackage[utf8]{inputenc}
\usepackage[english]{babel}
\usepackage{blindtext}
\usepackage{amsmath,amsfonts}
\usepackage{amssymb}
\usepackage{bbm}
\usepackage{algorithm}
\usepackage{algcompatible}
\usepackage[noend]{algpseudocode}
\usepackage{natbib}
\usepackage{setspace}
\usepackage{graphicx}
\usepackage{caption}
\usepackage{subcaption}
\usepackage{hyperref}
\hypersetup{
    colorlinks,
    allcolors=black
}
\usepackage{mathrsfs}
\usepackage{amsthm}
\usepackage{tikz}
\usepackage[section]{placeins}
\usepackage{float}
\usepackage{mathtools}
\usepackage{multirow}
\usepackage{xcolor}
\mathtoolsset{showonlyrefs}
\allowdisplaybreaks

\newcommand{\NN}{\mathbb{N}}

\newcommand{\RR}{\mathbb{R}}

\newcommand{\cdep}{\text{CDE}_P}

\newcommand{\wrt}[1]{\text{d}#1}

\newcommand{\Prb}[1]{\Pr\left(#1\right)}

\newcommand{\red}[1]{{\color{black}#1}}

\theoremstyle{definition}

\journal{Ocean Engineering}

\begin{document}

\begin{frontmatter}

\title{{Estimating Metocean Environments Associated with Extreme Structural Response to Demonstrate the Dangers of Environmental Contour Methods}}

\author[lancs]{Matthew Speers}
\author[shellnl]{David Randell}
\author[lancs]{Jonathan Tawn}
\author[lancs,shelluk]{Philip Jonathan}
\address[lancs]{Department of Mathematics and Statistics, Lancaster University LA1 4YF, United Kingdom.}	
\address[shellnl]{Shell Global Solutions International BV, Amsterdam, The Netherlands.}	
\address[shelluk]{Shell Research Limited, London, United Kingdom.}

\begin{abstract}
Extreme value analysis (EVA) uses data to estimate long-term extreme environmental conditions for variables such as significant wave height and period, for the design of marine structures. Together with models for the short-term evolution of the ocean environment and for wave-structure interaction, EVA provides a basis for full probabilistic design analysis. {Alternatively}, environmental contours provide an {approximate} approach to estimating structural integrity, without requiring structural knowledge. {These} contour methods also exploit statistical models, including EVA, but avoid the need for structural modelling by making what are believed to be conservative assumptions about the shape of the structural failure boundary in the environment space. These assumptions, however, may not always be appropriate, or may lead to unnecessary wasted resources from over design.  {We demonstrate a methodology for efficient fully probabilistic analysis of structural failure. From this, we estimate the joint conditional probability density of the environment ($\text{CDE}$), given the occurrence of an extreme structural response. We use $\text{CDE}$ as a diagnostic to highlight the deficiencies of environmental contour methods for design; none of the IFORM environmental contours considered characterise $\text{CDE}$ well for three example structures.}

\end{abstract}


\end{frontmatter}


\section{Introduction} \label{section:introduction}

\subsection{Background}

Ocean engineers use different approaches to quantify extreme conditions for design and reassessment of offshore and coastal structures. The natural full probabilistic approach (henceforth, the ``forward'' approach) is to construct a sequence of statistical models to characterise the extreme multivariate ocean environment, as well as the interaction between that environment and the structure (e.g., \citealt{towe2021efficient}). This approach considers the response of the structure to be a stochastic function of the environment, summarised by underlying sea state statistics such as significant wave height and period, contrary to previous work (e.g., \citealt{coles1994statistical}) where a deterministic relationship is assumed. The forward approach thus seeks a multivariate distribution $F_\mathbf{X}$ for environmental variables $\mathbf{X}$, as well as a distribution $F_{R_L|\mathbf{X}}$ which characterises the maximum stochastic response $R_L$ induced on the structure by the environment $\mathbf{X}$ over the period of a sea state of duration $L$ (e.g., 3) hours. A key property of this method is that uncertainty from the estimation of each distribution can be naturally quantified and propagated through the sequence of models. Structural risk assessment centres on the estimation of the probability of structural failure using the distribution $F_{R_S}$ for the maximum response $R_S$ in a random storm, or alternatively the distribution $F_{R_A}$ for the maximum response $R_A$ per annum, evaluated by marginalisation of the distribution $F_{R_L|\mathbf{X}}$ over the environment space. In addition, $F_{R_L|\mathbf{X}}$ may be used to find the probability of the event of structural failure $\mathcal{F}_L$ within a sea state (of duration $L$) with environmental conditions $\mathbf{X}$. 

{A combination of better models for the extreme ocean environment, techniques to reduce the computational complexity of the forward approach, and improved computational resources, have made forward estimation of structural failure probability more routinely achievable.} These include the development of practically-useful statistical models for non-stationary, or covariate effected, margins (e.g., \citealt{ChvDvs05,RndEA15a,Yng19a}) as well as the conditional multivariate extremes model developed by \cite{HffTwn04}, conditional simulation of extreme time series (of waves and wave kinematics) proposed by \cite{TylEA97}, and efficient importance sampling from distributions (e.g., \citealt{GlmEA04}). {In this work we demonstrate an efficient forward approach to estimate the tail of the distributions $F_{R_L|\mathbf{X}}$ and $F_{R_S}$, corresponding to return periods $P$ of the order of $10^3$ years. Specifically, we estimate the distribution of extreme base shear, approximated using the loading equation of \cite{MrsEA50} on simple structures, following the procedure of \cite{romans1995response}.} The environmental and structural models utilised here are sufficiently complex to illustrate key methodological steps, whilst being simple enough to avoid unnecessary complexity. We will also estimate the joint conditional density of environmental variables (henceforth referred to as $\text{CDE}_P$) given the occurrence of an extreme $P$-year Morison load on example structures. {$\text{CDE}_P$ is introduced as a diagnostic tool \emph{emerging from} fully probabilistic analysis, highlighting regions of environmental space associated with extreme structural responses. {We use $\text{CDE}_P$ to assess the relevance of other approaches which aim to identify regions of the environment space important to design.} 

Historically, forward estimation of the probability of failure has proved computationally intractable or prohibitively expensive. Instead, metocean design has tended to focus on a dominant variable (such as significant wave height) at a location, placing less emphasis on other associated environmental variables (e.g., \citealt{FldRndWuEA14}); the metocean engineer's challenge is then to estimate marginal return values of the dominant variable, and perhaps follow an engineering recipe to specify design values for associated variables. The development of methods of structural reliability, associated with \cite{MdsEA86}, made it clearer that good models for the joint distribution of environmental variables were necessary. For this reason, approaches to structural reliability which make use of more than one environmental variable became popular, such as environmental design contours derived from parametric hierarchical models of the environmental variables $\mathbf{X}$ (e.g., \citealt{Hvr87}). A recent review of statistical methodologies for metocean design is provided by \cite{VnmEA22}.

In one respect, contour methods are advantageous over the forward approach in that they characterise the environment only and so just require estimation of the joint environmental distribution $F_\mathbf{X}$. Therefore, combined with appropriate assumptions for the nature of the ocean-structure interaction, environmental contours can be used in principle to assess any structure in that environment. {The decoupling of environment and structure is achieved by making what are thought to be conservative assumptions about the environment-structure interaction, leading to what are believed to be conservative estimates of structural reliability. Many different methods exist to estimate environmental design contours (e.g., \citealt{Ross2020, HslEA21, mackay2021marginal, hafver2022environmental, mackay2023model}). These methods, whilst each attempting to construct the $P$-year level contour, can produce quite different estimates to one another. In addition, the environmental contour produced by a particular contour estimation method will vary with the estimate of $F_\mathbf{X}$ used to obtain it. In this work, we consider the IFORM environmental contour method (e.g., \citealt{HvrWnt09}), recommended by both \cite{standard2017actions} and \cite{dnv2017} standards. 
We focus on IFORM due to its popularity in the ocean engineering community. {We will assess the relative performance of different hierarchical models for $F_\mathbf{X}$ and their respective IFORM environmental contours, by quantifying their capacity to enclose $\text{CDE}_P$ for a given structure. We will show that, regardless of the hierarchical model form for $F_\mathbf{X}$, environmental contours do not provide reliable coverage of $\text{CDE}_P$ across a set of simple structure examples. Therefore, evaluation of response along these IFORM contour boundaries will not provide reliable realisations of the desired $P$-year response.
\subsection{Objectives and Layout}
The objective of the article is to promote the use of fully probabilistic design in favour of environmental contour-based methods. To achieve this, we provide the following analysis. (a) We demonstrate that fully probabilistic design (using the ``forward approach'') can be achieved in a computationally efficient manner and, motivated by this demonstration, recommend more routine adoption of fully probabilistic design. (b) Using the forward approach, we estimate the conditional density of the environment (CDE) across different example structures. We view CDE as a design diagnostic which identifies regions of the space of environmental variables contributing to extreme structural responses. We observe that this region changes from structure to structure. (c) We highlight the deficiencies of environmental contour methods for design, by assessing their ability to characterise the CDE. Specifically, we show that (c-i) a specified approach to environmental contour estimation is not suitable to characterise CDEs corresponding to different structures (because the environmental contour is structure-independent), and (c-ii) for reasonably-sized samples, the characteristics of the estimated environmental contour are sensitive to the modelling choices underpinning contour estimation, and that making these choices well is challenging. Our findings from (c) further motivate rejection of contour-based design in favour of the fully probabilistic alternative. \red{Further, if contour-based methods are to be used, they should be calibrated for the specific structural archetype under consideration. To establish this calibration, the full forward model would nevertheless need first to be evaluated for the archetype.}

The layout of the article is as follows. In Section~\ref{section:motivation}, we seek to motivate our analysis using a sample of data for storm peak significant wave height and second spectral moment wave period from a location in the central North Sea. Section~\ref{section:methodology} describes the methodologies combined to achieve the {efficient} forward approach for estimation of the distributions $F_{R_S}$ and $F_{R_A}$. The approach used to estimate environmental design contours, and the various parametric forms considered for the hierarchical estimation of $F_\mathbf{X}$ are given in Section~\ref{section:contours}. In Section~\ref{section:results}, we present estimates of $\cdep$ for three variants of a simple stick structure, and use these to quantify the performance of different IFORM environmental contours, {demonstrating that none of the IFORM contours estimates performs well for all example structures.} In Section~\ref{section:discussion}, we discuss the implications of our results, and make recommendations for structural design practice. We provide an online Supplementary Material (SM) with a fuller description of aspects of the procedures above.

\section{Motivating metocean dataset} \label{section:motivation}
We motivate the analysis using hindcast data for sea state significant wave height and second spectral moment wave period for a location in the central North Sea. The data consist of 124671 observations for the period January 1979 to September 2013, calculated for consecutive 3-hour sea states. Intervals corresponding to storm events are isolated from the hindcast data, using the approach of \cite{ewans2008effect}, resulting in a total of 2462 values for storm peak significant wave height ($H_S$) and corresponding wave period ($T_2$), for an average of $73$ storm events per annum. Figure~\ref{figure:hst2} shows the storm peak data $(H_S, T_2)$. Despite the fact that we expect these variables not to be identically distributed due to environmental covariates (e.g., direction and season), for the purposes of the current work we assume these to be independently and identically distributed.
\begin{figure}[h]
\begin{subfigure}[b]{0.49\textwidth}
    \centering
    \includegraphics[width=1\textwidth]{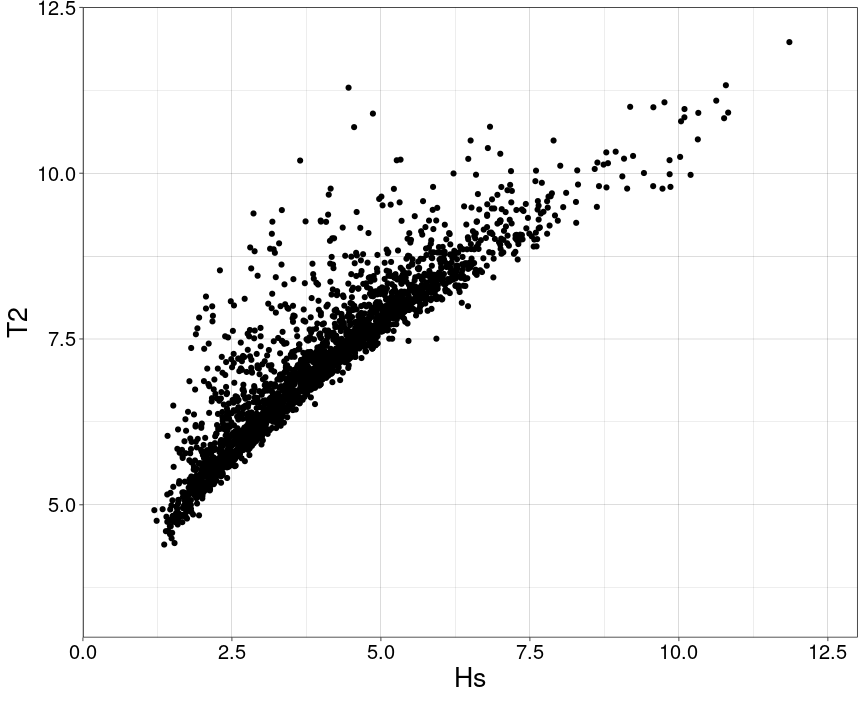}
    \caption{Storm peak variables $(H_S, T_2)$.}
    \label{figure:hst2}
\end{subfigure}
\hfill
\begin{subfigure}[b]{0.49\textwidth}
    \centering
    \includegraphics[width=1\textwidth]{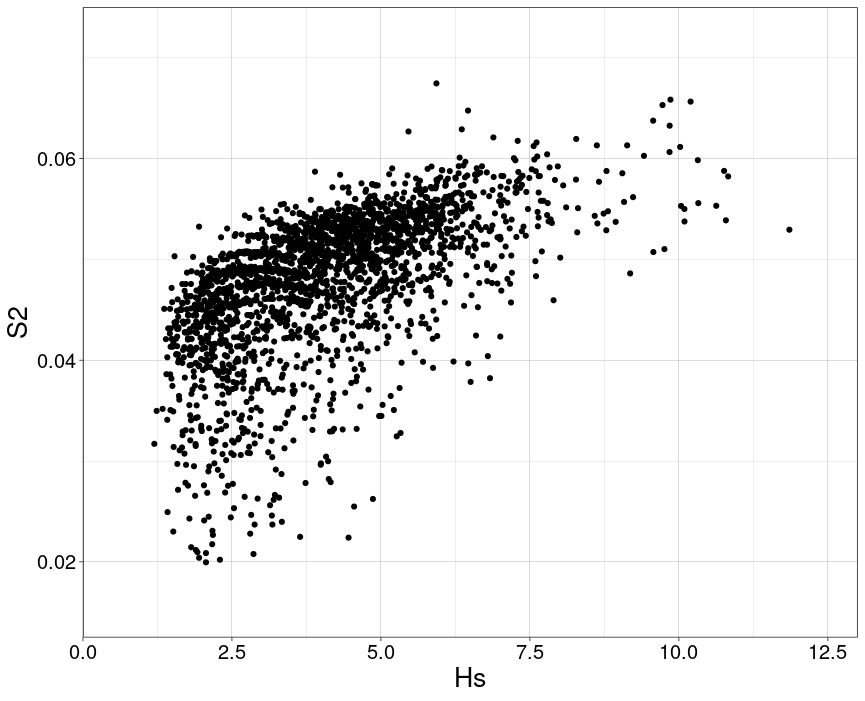}
    \caption{Storm peak variables $(H_S, S_2)$.} 
    \label{figure:hss2}
\end{subfigure}
\caption{Data of storm peak sea state variables from a location in the central North Sea. Storm peaks are extracted from hindcast data of consecutive $3$-hour sea states, using the method of \cite{ewans2008effect}, in terms of significant wave height $H_S$ [m], second spectral moment wave period $T_2$ [s] and wave steepness $S_2$.}
\label{figure:dataset}
\end{figure}
We choose to use storm peak wave steepness $S_2$ in favour of wave period $T_2$ in the analysis below. Values of $S_2$ are calculated via $S_2 = 2\pi H_S / (g T_2^2)$, where $g$ is the acceleration due to gravity. Figure~\ref{figure:hss2} shows the resulting storm peak data $\mathbf{X}=(H_S, S_2)$. Note throughout the paper, that we restrict the notation $H_S$, $S_2$ and $T_2$ to refer to storm peak quantities, and further that all numerical values are quoted in SI units. Additionally, all physical properties are taken to be real-valued unless stated otherwise.

The choice of $(H_S, S_2)$ over $(H_S, T_2)$ is motivated by the fact that extreme value models are generally developed for the joint upper tail of variables. Extreme environmental loads are typically generated by large values of $H_S$, large values of $S_2$ but non-extreme values of $T_2$; that is, joint extremes of $(H_S, S_2)$ in the upper tail induce the most extreme environments and structural responses. Therefore, it is appealing to structure the analysis in terms of $(H_S, S_2)$ (e.g., \citealt{myrhaug2018some}). 


\section{Environment and response modelling} \label{section:methodology}

\subsection{Outline of the forward approach and $\text{CDE}_P$} \label{section:methodology:mathsOutline}

\subsubsection{Storm peak characteristics and intra-storm evolution} 
We illustrate the methodologies combined to form the forward approach for direct estimation of the distribution of extreme structural response (specifically, base shear), in order to estimate the conditional distribution of the environment $\text{CDE}_P$. For generality, and alignment with the work of others (e.g., \citealt{towe2021efficient}), initially we present a form of the forward approach which incorporates full intra-storm evolution and the effects of covariates. We subsequently restrict the model, and focus our analysis on storm peak variables as introduced in Section~\ref{section:motivation}, so as to emphasise the key methodological steps only. 

We assume access to metocean data for storm peak variables $\mathbf{X}^\text{sp}$ (e.g., $H_S$ and $S_2$ from Section~\ref{section:motivation}). Since the characteristics of storm peak events generally vary with respect to covariates (e.g., \citealt{randell2015distributions}) we also assume access to storm peak covariates $\boldsymbol{\Theta}^\text{sp}$, which encode this information. The response of an offshore structure to environmental loading occurs continuously, and in particular for the full duration of the storm event, consisting of a series of sea states (of duration $L$ hours) indexed by $ s \in \mathcal{S}_T = \{1, 2,\ldots, T\}$, for unknown storm length $T$. To estimate the distribution of the maximum structural response $R_S$ in a storm, we need to consider (a) the variability of the duration $T$ of a random storm, (b) the full evolution of sea state variables $\{\mathbf{X}_s\}_{s \in \mathcal{S}_T}$ over a given storm (themselves not identically distributed given the sea state covariates $\{\boldsymbol{\Theta}_s\}_{s \in \mathcal{S}_T}$) and (c) the resulting maximum response $R_S|\{(\mathbf{X}_s, \boldsymbol{\Theta}_s)\}_{s \in S_T}$ induced over sea states within the storm.

As sea states are dependent over time, $R_{L,s}$ and $R_{L,s'}$, for any $s,s' \in \mathcal{S}_T$ where $s\neq s'$, will also be dependent, where $R_{L, s}$ is the maximum response in a sea state of length $L$ at time index $s$. However, this dependence is purely due to the sea state characteristics $\{(\mathbf{X}_s, \boldsymbol{\Theta}_s)\}_{s \in\mathcal{S}_T}$ evolving over time.  This is because a load $R_{L,s}$ from the model of \cite{MrsEA50} is produced by an individual random wave event, and the statistical properties of wave events within a sea state at time index $s$ are determined by the characteristics $(\mathbf{X}_s, \boldsymbol{\Theta}_s)$, yet the interval of time over which consecutive waves are correlated is considerably shorter than the length $L$ of a sea state. Therefore, the conditional dependence between the variables $R_{L,s}$ and $R_{L,s'}$, given $(\mathbf{X}_s, \boldsymbol{\Theta}_s)$ and $(\mathbf{X}_{s'}, \boldsymbol{\Theta}_{s'})$, is negligible. We exploit this reasoning to facilitate step (c), where we assume that the random variables $\{R_{L,{s}}\}_{s\in \mathcal{S}_T}$ are conditionally independent given sea states $\{(\mathbf{X}_{s}, \boldsymbol{\Theta}_{s})\}_{s \in \mathcal{S}_T}$, which leads to the following simplification, for response $r>0$
\begin{align}\label{equation:conditional_assumption}
    F_{R_S|\{(\mathbf{X}_s, \boldsymbol{\Theta}_s)\}_{s \in S_T}}(r|\{(\mathbf{x}_s, \boldsymbol{\theta}_s)\}_{s \in S_\tau}) = \prod_{s \in \mathcal{S}_T} F_{R_{L}|(\mathbf{X}_s, \boldsymbol{\Theta}_s)} (r|\mathbf{x}_s, \boldsymbol{\theta}_s),
\end{align}
where we omit the subscript $s$ in writing $F_{R_L|(\mathbf{X}_s, \boldsymbol{\Theta}_s)}$ since the dependence of $R_L$ on $s$ is contained through the vales of $(\mathbf{X}_s, \boldsymbol{\Theta}_s)$.

\subsubsection{Distribution of maximum response per storm and per annum} \label{section:storm_max_defninition}

The forward approach estimates the cumulative distribution function $F_{R_S}$ of the maximum structural response $R_S$ in a storm, and subsequently the distribution of maximum response $R_A$ per annum. Estimation of $F_{R_S}$ requires (a) modelling of multivariate storm peak variables $\mathbf{X}^\text{sp}$ given storm peak covariates $\boldsymbol{\Theta}^\text{sp}$, (b) characterisation of the conditional time-varying within-storm evolution of sea state characteristics $\{(\mathbf{X}_s,\boldsymbol{\Theta}_s)\}_{s \in \mathcal{S}_T}$ given storm peak characteristics $\mathbf{X}^\text{sp}, \boldsymbol{\Theta}^{\text{sp}}$, and (c) the estimation of the maximum response $R_{L, s}$ given sea state characteristics $(\mathbf{X}_s, \boldsymbol{\Theta}_s)$ for an $L$-hour sea state. We must also consider the variability in the duration $T$ of a storm event. See \cite{towe2021efficient} for previous discussions of similar models. Given knowledge of the above, and exploiting assumption \eqref{equation:conditional_assumption} made in Section~\ref{section:methodology:mathsOutline}, the distribution of $R_S$ can be written as 
\begin{align}
F_{R_S}(r) &=  \int_{\boldsymbol{\theta}^\text{sp}} \int_{\mathbf{x}^\text{sp}} \int_{(\left\{(\mathbf{x}_s, \boldsymbol{\theta}_s\right)\}_{s \in \mathcal{S}_\tau}, \tau)} \prod_{s \in \mathcal{S}_\tau} F_{R_{L}|(\mathbf{X}_s, \boldsymbol{\Theta}_s)} (r|\mathbf{x}_s, \boldsymbol{\theta}_s)\\ & \times f_{(\left\{(\mathbf{X}_s, \boldsymbol{\Theta}_s)\right\}_{s \in \mathcal{S}_T}, T) | \mathbf{X}^\text{sp}, \boldsymbol{\Theta}^\text{sp}}\left(\left\{(\mathbf{x}_s, \boldsymbol{\theta}_s)\right\}_{s \in \mathcal{S}_\tau}, \tau \mid \mathbf{x}^\text{sp}, \boldsymbol{\theta}^\text{sp}\right) \\ & \times f_{\mathbf{X}^\text{sp}|\boldsymbol{\Theta}^\text{sp}}(\mathbf{x}^\text{sp}|\boldsymbol{\theta}^\text{sp}) \times f_{\mathbf{\Theta}^\text{sp}} (\boldsymbol{\theta}^\text{sp}) \,\, \wrt{(\{(\mathbf{x}_s, \boldsymbol{\theta}_s)\}_{s \in \mathcal{S}_\tau}, \tau)} \,\,\wrt{\mathbf{x}^\text{sp}}  \,\,\wrt{\boldsymbol{\theta}^\text{sp}},
\label{eqn:forwardModel}
\end{align}
for $r >0$, where $f_{\boldsymbol{\Theta}^\text{sp}}$ is the joint probability density of storm peak covariates, $f_{\mathbf{X}^\text{sp}|\boldsymbol{\Theta}^\text{sp}}$ is the joint probability density of storm peak variables given storm peak covariates, and $f_{(\left\{(\mathbf{X}_s, \boldsymbol{\Theta}_s)\right\}_{s\in \mathcal{S}_T}, T )| \mathbf{X}^\text{sp}, \boldsymbol{\Theta}^\text{sp}}$ is the joint probability density for the full time-series evolution of within-storm sea state characteristics (and storm duration) given storm peak characteristics. 

If we assume that the number of occurrences of storm events in a year is Poisson-distributed with expectation $\lambda$ per annum, we can use $F_{R_S}$ to estimate the corresponding cumulative distribution function $F_{R_A}$ of the maximum response $R_A$ in a year, i.e., 
\begin{equation}
F_{R_A}(r)= \sum_{m=0}^\infty [F_{R_S}(r)]^m \frac{\lambda^m e^{-\lambda}}{m!} = \exp \left[-\lambda\left(1-F_{R_S}(r)\right)\right].
\end{equation}
for $r  >0$. From this expression, we may define return values for maximum response corresponding to a return period of $P$ years as $r_P = F_{R_A}^{-1}(1-1/P)$.

\subsubsection{Reduced forward approach} \label{section:reduced}

The objective of the current work is to compare estimates for $\cdep$ made using the forward approach, summarised by \eqref{eqn:forwardModel}, with environmental contours at the $P$-year level. This comparison is useful even in the absence of covariate effects, and whilst neglecting intra-storm evolution. Therefore, to minimise computational complexity, we now assume that the effects of covariates $\boldsymbol{\Theta}^\text{sp}$ and $\boldsymbol{\Theta}_s$ can be ignored, and that the maximum response in a storm always occurs in the storm peak sea state, so that intra-storm evolution can also be ignored. As a result, integral \eqref{eqn:forwardModel} for $F_{R_S}$ reduces to 
\begin{equation} \label{equation:simplified_old_notation}
    F_{R_S}(r) = \int_{\mathbf{x}^\text{sp}} F_{R_{L}|\mathbf{X}^\text{sp}}(r| \mathbf{x}^\text{sp}) f_{\mathbf{X}^\text{sp}}(\mathbf{x}^\text{sp})  \wrt{\mathbf{x}^\text{sp}},
\end{equation} 
for $r >0$, where $F_{R_{L|\mathbf{X}^\text{sp}}}$ is the cumulative distribution function of the maximum response over an $L$-hour sea state given storm peak variables $\mathbf{X}^\text{sp}$. For brevity, we henceforth omit the storm peak superscript and write
\begin{equation}
F_{R_S}(r) = \int_{\mathbf{x}} F_{R_{L}|\mathbf{X}}(r| \mathbf{x}) f_{\mathbf{X}}(\mathbf{x})  \wrt{\mathbf{x}},
\label{eqn:forwardModelSimple}
\end{equation}
for $r  >0$, where $f_{\mathbf{X}}$ is now the joint density of storm peak variables $\mathbf{X}=(X_1,\ldots,X_p)$, and $F_{R_{L}|\mathbf{X}}$ is the distribution of maximum response over an $L$-hour storm peak sea state with variables $\mathbf{X}$. This brings us back to the setting of Section~\ref{section:motivation}, where we introduced observations of storm peak sea state data $\mathbf{X}$ with $p=2$. The key steps in evaluating the reduced forward approach in \eqref{eqn:forwardModelSimple} therefore become the estimation of $f_{\mathbf{X}}$ and $F_{R_{L}|\mathbf{X}}$. The first of these inferences is achieved using the conditional extremes model of \cite{HffTwn04}, as described in Section~\ref{section:conditional_extremes}. The second inference involves conditional simulation of environmental time-series following \cite{TylEA97}, and importance sampling from the tail of $F_{R_{L}|\mathbf{X}}$ following \cite{towe2021efficient}, as described in Section~\ref{section:response_dist}. 

\subsubsection{Probability of structural failure} \label{section:probability_failure}

When designing offshore structures, it is often desirable to determine the probability of the event of structural failure $\mathcal{F}_L$ for a given $L$-hour sea state with variables $\mathbf{X}=\mathbf{x}$. 
Given an estimate for the distribution of $R_L|(\mathbf{X}=\mathbf{x})$, we can evaluate this probability using the expression
\begin{equation}\label{equation:failure_probability}
    \Prb{\mathcal{F}_L|\mathbf{X}=\mathbf{x}} = \int_{r_L} \Prb{\mathcal{F}_L|R_L=r_L} f_{R_L|\mathbf{X}} (r_L |\mathbf{x}) \, \wrt{r_L},
\end{equation}
where the failure probability $\Prb{\mathcal{F}_L|R_L=r_L}$ depends on the nature of the structure, as illustrated in Figure~\ref{fig:fail_probs_trio} in Section~\ref{section:results}.

\subsubsection{Conditional density of the environment} \label{section:condition_density_definition}

The conditional density of the environment $f_{\mathbf{X}|R_L}(\cdot|r_P)$ describes the joint density of the environmental variables $\mathbf{X}$, conditional on the appearance of a $P$-year maximum response $r_P$ within a sea state of length $L$ hours. Given estimates for $f_{\mathbf{X}}$, $F_{R_{L}|\mathbf{X}}$, $F_{R_L}$ and $r_P$,  $\text{CDE}_P$ can therefore be evaluated using Bayes' rule 
\begin{equation}\label{equation:CDE}
f_{\mathbf{X}|R_L}(\mathbf{x}|r_P) = \frac{f_{R_L|\mathbf{X}}(r_P|\mathbf{x}) f_\mathbf{X}(\mathbf{x})}{f_{R_L}(r_P)}, 
\end{equation}
for $\mathbf{x} \in \RR^p$, where $f_{R_L|\mathbf{X}}$ and $f_{R_L}$ are the densities corresponding to the distributions $F_{R_L|\mathbf{X}}$ and $F_{R_L}$ respectively. Examples of $\cdep$ for the central North Sea application are given in Section~\ref{section:results_CDE}.

\subsection{Joint modelling of storm peak conditions} \label{section:conditional_extremes}

\subsubsection{Outline of the conditional extremes model}
The upper extremes of the marginal and joint distributions of the environmental variables $\mathbf{X}=(X_1,\ldots,X_p)$, corresponding to the storm peak sea state, are described using the conditional extremes model of \cite{HffTwn04}. For our illustrative example, in Section~\ref{section:motivation}, $p=2$ but we present the methodology here and for environmental contours for dimension $p$, to cover more general cases. This asymptotically justified flexible framework allows for the characterisation of joint tail behaviour from a sample of independently identically distributed observations of $\mathbf{X}$, without the need for making a subjective choice for a particular form of extremal dependence model (copula) between variables. This method has been applied extensively to the modelling of oceanographic data (e.g., \citealt{jonathan2014non, Towe2019model, shooter2021basin,tendijck2023modeling}).

The conditional extremes method uses univariate extreme value techniques to characterise the distribution of each variable individually, with the joint structure specified for variables on standard (typically Laplace) marginal scales (e.g., \citealt{Keef2013}). Estimation of the conditional extremes model is thus performed in two stages: (a) marginal extreme value modelling of each variable $X_j$ ($j=1,\ldots, p$) in turn, followed by the marginal transformation $X_j \mapsto Y_j$ of each variable to standard Laplace scale $Y_j$ and (b) estimation of the conditional extremes model for the set of Laplace-scale variables $\mathbf{Y}=(Y_1, \ldots, Y_p)$. Subsequently, we estimate $f_\mathbf{X}$ within an extreme joint tail region using the fitted conditional extremes model. The above steps are discussed below. 

\subsubsection{Marginal modelling and marginal transformation to Laplace scale} \label{section:conditional_extremes:marginal} 

We adopt the approach of \cite{davison1990models} for marginal modelling of storm peak variable $X_j$ for $j=1,\ldots,p$. We fit a generalised Pareto distribution (GPD) to exceedances of high threshold $u_j$, and model threshold non-exceedances empirically. Our marginal model $F_{X_j}$ for the cumulative distribution function of $X_j$ can thus be written
\begin{equation} \label{equation:marginal_model}
F_{X_j}(x) = 
\begin{cases}
    \Tilde{F}_{X_j}(x) & x \leq u_j \\
    \Tilde{F}_{X_j}(u_j) + \{1-\Tilde{F}_{X_j}(u_j)\} F_{\text{GPD},j}(x; u_j, \sigma_j, \xi_j) & x > u_j,
\end{cases}
\end{equation}
where $\Tilde{F}_{X_j}$ is the empirical distribution of $X_j$ and
$$
F_{\text{GPD},j}(x;u_j, \sigma_j, \xi_j) = 1 - \left( 1+ \frac{\xi_j (x-u_j)}{\sigma_j}\right)_+^{-1/\xi_j},
$$
for $x>u_j$, with scale and shape parameters $\sigma_j>0$ and $\xi_j \in \mathbb{R}$ and where $y_+ = \max(y, 0)$ for $y \in \RR$. The values of $u_j$ ($j=1,\ldots,p$) are selected using the univariate extreme threshold selection methods summarised by \cite{coles2001introduction}, see SM. Parameters $\sigma_j$ and $\xi_j$ are jointly estimated using standard maximum likelihood techniques. The probability integral transform 
\begin{equation} \label{equation:prob_int_transform}
Y_j= \begin{cases}\log \left\{2 F_{X_j}\left(X_j\right)\right\} & \text { for } X_j<F_{X_j}^{-1}(0.5) \\ -\log \left\{2\left[1-F_{X_j}\left(X_j\right)\right]\right\} & \text { for } X_j > F_{X_j}^{-1}(0.5),\end{cases}
\end{equation}
is then applied to each variable in turn, obtaining the standard Laplace-scaled equivalents $\mathbf{Y}$ of $\mathbf{X}$, such that, for $j=1,\ldots, p$, the distribution of $Y_j$ is
$$
F_{Y_j}(y) = 
\begin{cases}\frac{1}{2} \exp \left({y}\right) &  y \leq 0 \\ 1-\frac{1}{2} \exp \left(-{y}\right) &  y > 0.\end{cases}
$$

\subsubsection{Joint dependence modelling} \label{section:conditional_extremes:joint}

Having transformed environmental variables $\mathbf{X}$ to standard Laplace-scale equivalents $\mathbf{Y}$, we now apply the model of \cite{HffTwn04} to estimate the joint distribution $F_{\mathbf{Y}}$ in the upper tail region. It is shown by \cite{HffTwn04} and \cite{Keef2013} that, for $j=1,\ldots, p$, there exist unique values of parameters $\boldsymbol{\alpha}_{|j} \in [-1, 1]^{p-1}$, $ \boldsymbol{\beta}_{|j}\in (-\infty, 1]^{p-1}$, satisfying the constraints of \cite{Keef2013}, and $\mathbf{z}_{|j} \in \RR^{p-1}$, $y>0$, such that
\begin{equation} \label{equation:cond_extremes_limit}
    \lim_{v_j\rightarrow \infty }\Prb{\frac{\mathbf{Y}_{-j} -\boldsymbol{\alpha}_{|j} Y_j}{Y_j^{\boldsymbol{\beta}_{|j}}} < \mathbf{z}_{|j}, Y_j - v_j >y | Y_j >v_j } = e^{-y} G_{|j}(\mathbf{z}_{|j}),
\end{equation}
where the $(p{-}1)$-vector $\mathbf{Y}_{-j}$ denotes the $p$-vector $\mathbf{Y}$ with $j$th element $Y_j$ removed, $G_{|j}$ is a $(p{-}1)$-dimensional distribution function with non-degenerate marginals, and componentwise operations are assumed. Property \eqref{equation:cond_extremes_limit} can be leveraged by assuming a non-linear regression of $\mathbf{Y}_{-j}$ onto $Y_j$ holds for all values of $\mathbf{Y}$ within the region $\{\mathbf{Y} \in \RR^p : Y_j > v_j\}$, for some suitably large finite threshold $v_j>0$. For conditioning variable $Y_j$ ($j=1,\ldots, p$), the form of this regression is
\begin{equation}
\label{equation:ht_model}
(\mathbf{Y}_{-j}|\{Y_j=y\}) = \boldsymbol{\alpha}_{|j} y + y^{\boldsymbol{\beta}_{|j}}\mathbf{Z}_{|j} , \quad y > v_j,
\end{equation}
where $\mathbf{Z}_{|j}\sim G_{|j}$ is a $(p{-}1)$-dimensional residual random variable that is independent of $Y_j$ given $Y_j>v_j$. We estimate parameter vectors $\boldsymbol{\alpha}_{|j}$ and $\boldsymbol{\beta}_{|j}$ using standard maximum likelihood techniques, assuming for model fitting only that $G_{|j}$ corresponds to independent Gaussian distributions with unknown means and variances. The distribution $G_{|j}$ is modelled via the kernel density estimate of the observed values of the $(p{-}1)$-dimensional residual
\begin{equation}\label{equation:ht_residuals}
    \mathbf{Z}_{|j} = \frac{\mathbf{Y}_{-j} - \boldsymbol{\alpha}_{|j} Y_j}{Y_j^{\boldsymbol{\beta}_{|j}}}, \quad \text{for } \, Y_j > v_j,
\end{equation}
as in \cite{winter2017kth}.

\subsubsection{Simulation under the conditional model and estimation of the environment joint density}\label{section:conditional_extremes:density}

Inferences in $\RR^p$ using the fitted conditional extremes model are typically made by careful combination of Laplace-scale simulations in each of the upper tail regions $\{\mathbf{Y} \in \RR^p : Y_j > v_j\}$, for $j=1,\ldots,p$, together with empirical estimation in the remaining region $\{\mathbf{Y} \in \RR^p : Y_j \leq v_j \, \forall j\}$, as described in \cite{HffTwn04}, to give a set of size $N_{\text{sim}}$ realisations from the estimate of the joint density ${f}_\mathbf{\mathbf{Y}}$. The $p$ fitted marginal models \eqref{equation:marginal_model} can then be used componentwise to transform this sample of $\mathbf{Y}$ with Laplace-scale marginals to a sample of $\mathbf{X}$ on the original physical scale.

We can use the simulated sample to estimate the probability of $\mathbf{X}$ being in sub-regions of $\RR^p$. Specifically, if $D$ is the set of feasible $\mathbf{X}$ values such that $\Prb{\mathbf{X} \in \RR^p\setminus D}=0$, then we consider a partition $(D_1, \ldots, D_M)$ of $D$. Then, if $N_{\text{sim},i}$ is the number of realisations in set $D_i$, we can estimate $\Prb{\mathbf{X} \in D_i}$, for any $i=1, \ldots ,M$, as ${\Pr}({\mathbf{X} \in D_i})={N_{\text{sim},i}}/N_{\text{sim}}$. To obtain an estimate $f_{\mathbf{X}}(\mathbf{x})$ for any $\mathbf{x}\in D$ we exploit the property that, if $\mathbf{x}\in D_i$ and $|D_i|$ is sufficiently small that ${f}_{\mathbf{X}}$ is reasonably constant for all $\mathbf{x}\in D_i$, then
\begin{equation}\label{equation:density_approximation}
    {\Pr}({\mathbf{X} \in D_i}) =\int_{\mathbf{x}'\in D_i} {f}_\mathbf{X}(\mathbf{x}') \wrt{\mathbf{x}'}\approx |D_i|{f}_\mathbf{X}(\mathbf{x}),
\end{equation}
yielding the estimate $\hat{f}_\mathbf{X}(\mathbf{x}) = N_{\text{sim, i}}/(N_\text{sim}|D_i|)$ for all $\mathbf{x} \in D_i$. We can achieve the required conditions for the approximation to be reliable by taking $M$ to be sufficiently large and selecting all $D_i$ such that $|D_i|\propto M^{-1}$. 

\subsection{Estimation of maximum response in a storm peak sea state given storm peak variables} \label{section:response_dist}

\subsubsection{Outline of estimation of $F_{R_L|\mathbf{X}}$}

To derive properties of $R_L$ we first need to model the behaviour of the maximum response $R_I$ to an individual wave in sea state $\mathbf{X}$. However, to estimate the distribution $F_{R_I|\mathbf{X}}$ of the maximum response due to the action of an individual wave in sea state $\mathbf{X}$, we first evaluate the distribution $F_{R_I|(\mathbf{X}, C)}$ of the maximum response $R_I$ to an individual wave in the sea state $\mathbf{X}$ with crest elevation $C$. This is achieved by simulation of wave fields under sea state conditions $\mathbf{X}$ with known crest elevation $C$, followed by propagation of the resulting stochastic wave fields through to the structural response model; see Section~\ref{section:conditional_wave_simulation} for details. Then, integrating out $C$, we have
\begin{equation} 
    F_{R_I|\mathbf{X}}(r|\mathbf{x}) = \int_{\RR^+} F_{R_I|\mathbf{X}, C}(r|\mathbf{x}, c) f_{C|\mathbf{X}}(c|\mathbf{x}) \wrt{c},
    \label{equation:importance_sampling_target}
\end{equation}
for $r  >0$, where $f_{C|\mathbf{X}}$ is the density of crest elevation in the sea state $\mathbf{X}$, where we assume that crests are Rayleigh-distributed, with density
\begin{equation}\label{equation:rayleigh_density_0}
    f_{C|\mathbf{X}}(c|\mathbf{x}) = \frac{16c}{h(\mathbf{x})^2} \exp\left(-8\frac{c^2}{h(\mathbf{x})^2}\right).
\end{equation}
for $c>0$, with sea state  $\mathbf{X}=\mathbf{x}$ with significant wave height $h(\mathbf{x})=x_1$. Computationally efficient estimation of $F_{R_I|\mathbf{X}}$ following \eqref{equation:importance_sampling_target} is achieved using importance sampling; see Section~\ref{section:importance_samp}.  Finally, we obtain $F_{R_L|\mathbf{X}}$ from $F_{R_I|\mathbf{X}}$ by assuming that (a) there are a fixed number of waves per $L$-hour sea state $\mathbf{x}$, given by $Q_L(\mathbf{x}) =60^2L/ t_2(\mathbf{x})$ where $t_2(\mathbf{x})$ is the second spectral moment wave period for the sea state, and (b) individual-wave maximum responses (i.e., the $R_I$) in a given sea state are independent of each other. Assumption (a) approximates the stochastic number of waves per sea state with an `average' value, and (b) holds since individual base shears calculated in Section~\ref{section:conditional_wave_simulation} are observed for fractions of a second, significantly less than the typical wave period; therefore, there is no correlation between responses induced by different waves for a known sea state. Combining these assumptions gives 
\begin{equation} \label{equation:powering_up_integral}
    F_{R_L|\mathbf{X}}(r|\mathbf{x}) = \{F_{R_I|\mathbf{X}}(r|\mathbf{x})\}^{Q_L(\mathbf{x})},
\end{equation}
for $r  >0$, following the definition for the distribution of the maximum of independent random variables.
 
\subsubsection{Simulation of maximum response to the action of an individual wave, given sea state variables and crest elevation} \label{section:conditional_wave_simulation}

We estimate the distribution of $R_I|(\mathbf{X}, C)$ in two stages: (a) simulation of realisations of wave fields under sea state conditions $\mathbf{X}$ with known crest elevation $C$, followed by (b) propagation of the resulting wave fields through a suitable structural response model. The details of each stage are outlined below.

The model of \cite{TylEA97} allows for conditional simulation of a wave field given the occurrence of a turning point of surface elevation in time, with specified crest elevation $C=c>0$ at the structure location at time $t=0$, for a given sea state $\mathbf{X}=\mathbf{x}$ with wave spectrum $S(\cdot; \mathbf{x})$, using linear wave theory. The JONSWAP spectrum of \cite{hasselmann1973measurements} is chosen as the form of $S(\cdot; \mathbf{x})$ due to its applicability to the North Sea wave conditions \citep{holthuijsen2010waves}, see SM for further details. \cite{TylEA97} provides expressions for linear crest elevation $E(t;\mathbf{x}, c)$, horizontal velocity $U(t; z, \mathbf{x}, c)$, and horizontal acceleration $\dot{U}(t; z, \mathbf{x}, c)$, at time $t\in\RR$ and vertical position $z\in \RR$, relative to the mean water level, each conditioned on the wave process (a) attaining a turning point of $E$ at time $t=0$, with (b) $E(t=0; \mathbf{x}, c)=c$, both at the location of the structure. The forms of $E(t; \mathbf{x}, c)$ and $U(t;z, \mathbf{x}, c)$ are
\begin{align} \label{equation:wave_elevation}
 E(t; \mathbf{x}, c)&=\sum_{n=1}^N \left\{(A_n + \mathcal{Q}\sigma^2_n) \cos \left(\omega_n t \right)+(B_n + \mathcal{R} \sigma^2_n \omega_n)\sin \left(\omega_n t \right)\right\},
\end{align}
and 
\begin{equation} \label{equation:wave_velocity}
U(t; z, \mathbf{x}, c) =
            \sum_{n=1}^N  \omega_n \frac{\cosh\left(k_n(d+z)\right)}{\sinh(k_nd)} \left\{(A_n + \mathcal{Q}\sigma^2_n) \cos(\omega_n t ) + (B_n + \mathcal{R} \sigma^2_n \omega_n) \sin(\omega_n t)\right\},
\end{equation}
for $z <  E(t; \mathbf{x}, c)$ and zero otherwise, for a regular grid of angular frequencies $\omega_1,\ldots, \omega_N>0$ with spacing $\delta\omega >0 $ and $N\in \NN$ specified below, where $A_n,B_n\in \RR$ ($n=1,\ldots, N)$ and $\mathcal{Q}, \mathcal{R}\in \RR$ are random coefficients, $
d$ is water depth and $k_n$ ($n=1,\ldots, N$) are wave numbers given implicitly by $\omega_n^2 = g k_n \tanh (k_n d)$. The equation for $\dot{U}(t;z, \mathbf{x}, c)$ can be found by differentiation of $U(t;z, \mathbf{x}, c)$ with respect to $t$. Coefficients $A_n,B_n$ $(n=1,\ldots, N)$ are a series of independently and identically distributed $N(0, \sigma^2_n)$ random variables with variance $\sigma^2_n= S(\omega_n;\mathbf{x})\delta\omega$, the integrated spectral density in the frequency band $(\omega_n - \delta\omega/2, \omega_n + \delta\omega/2)$ of the discretised wave spectrum. The random coefficients $\mathcal{Q}$ and $\mathcal{R}$ are defined as
$$
\mathcal{Q} =\frac{1}{\sum_n \sigma^2_n} \left( c - \sum_{n=1}^N A_n \right)\,\, \text{and} \,\, \mathcal{R} = \frac{1}{\sum_n \omega_n^2  \sigma^2_n} \left( - \sum^N_{n=1} \omega_n B_n\right).
$$
Next, we estimate the total base shear response of the structure to the simulated conditional wave field.  We assume the wave-structure interaction to be quantified by the equation of \cite{MrsEA50}, which estimates drag and inertial loads applied by the ocean environment on a stick structure. These loads are calculated from the wave velocity and acceleration fields respectively. Under the assumptions of linear wave theory, these fields can be derived entirely from knowledge of the wave spectrum. For the applications described in this work, we assume that the values of sea state $(H_S,S_2)$ are sufficient to define the wave spectrum; hence, for a given structure, the 2-dimensional storm peak representation $\mathbf{X}=(X_1,X_2)=(H_S,S_2)$ from Section~\ref{section:motivation} is sufficient to describe the extreme ocean environment and its associated structural response.

Under our simplifying assumptions, waves are assumed to be unidirectional, propagating in a single direction towards the vertical cylindrical structure with nominal small diameter. Waves are assumed to pass through the structure, whilst also exerting force, without being obstructed, and the effects of current and wind are ignored. This wave field model provides a basis to approximate the induced load on a jacket structure. The Morison loading equation estimates the base shear $M(t;z, \mathbf{x}, c)$ induced on a cylinder by the wave at time $t$ and vertical position $z$, and is given by
\begin{equation}\label{equation:morison_load}
    M(t;z, \mathbf{x}, c) =\rho c_m(z) V \dot{U}(t;z, \mathbf{x}, c)+\frac{1}{2} \rho c_d(z) A U(t;z, \mathbf{x}, c)|U(t;z, \mathbf{x}, c)|,
\end{equation}
where $c_m(z), c_d(z)>0$ are inertia and drag coefficients, $\rho=1024$ (recall that SI units are used throughout) is the density of water, $V$ is the volume of the body and $A$ is the area of the structure perpendicular to the wave propagation. We assume a cylindrical structure with diameter 1 and height of 150 situated within water of depth $d=100$. Since the probability of a crest elevation greater than 50 is near zero for all relevant sea states, this structure scenario amounts to a cylinder of infinite height. In order to approximate models of different structure types, $c_m(z)$ and $c_d(z)$ can be made to vary with $z$, as discussed in Section~\ref{section:results}. To evaluate the total base shear $B_S(t; \mathbf{x}, c)$ on the structure at time $t$, we integrate $M(t;z, \mathbf{x}, c)$ to give
\begin{equation}\label{equation:force_integral}
   B_S(t; \mathbf{x}, c) = \int_{-d}^{E(t; \mathbf{x}, c)} M(t;z, \mathbf{x}, c) \wrt{z},
\end{equation}
the total Morison load induced up the water column at the structure location.

The response $R_I|(\mathbf{X}=\mathbf{x}, C=c)$ may be obtained by considering the portion of the time series $\{B_S(t;\mathbf{x}, c)\}_{t \in \RR}$ that corresponds to the central wave conditioned to attain $E(t=0; \mathbf{x},c)=c$; that is, the period of time $t \in \mathcal{T}_0\subset \RR$, with $0 \in \mathcal{T}_0$, for which the wave surrounding the conditioning crest of elevation $c$ at time $t=0$ acts on the structure. We define
\begin{equation}\label{equation:per_wave_max}
R_I|(\mathbf{X}=\mathbf{x},C=c) = \max_{t \in \mathcal{T}_0}{B_S(t;\mathbf{x}, c)}.
\end{equation}
We obtain realisations of $R_I|(\mathbf{X}=\mathbf{x}, C=c)$ from a time series of the base shear response \eqref{equation:force_integral} evaluated using Morison loads \eqref{equation:morison_load}, in turn calculated from wave fields simulated according to expressions \eqref{equation:wave_elevation} and \eqref{equation:wave_velocity}. The interval of time over which wave fields are simulated corresponds to a period of $120$ seconds, sufficiently large to ensure reliable performance of the FFT algorithm \citep{cooley1965algorithm}, meaning here $\mathcal{T}_0 \subset [-60, 60]$. Realisations of conditional crest elevation and wave kinematics are simulated for a regular grid $\{(t_i,z_j)\}_{i=1,j=1}^{n_t,n_z}$ of values $t \in [-60, 60]$ and $z \in [-100, 150]$. We set $N=n_t$ in expressions \eqref{equation:wave_elevation} and \eqref{equation:wave_velocity}, which is necessary to evaluate the wave field equations using the FFT algorithm; see SM for details. The values of $n_t=480$ and $n_z=50$ are chosen, sufficient to ensure reasonable response approximation. The simulated kinematics are then propagated through the Morison equation, providing a realisation of $\{M(t_i; z_j, \mathbf{x}, c)\}_{i=1,j=1}^{n_t,n_z}$. Numerical evaluation of integral \eqref{equation:force_integral} with respect to $z$ yields a realisation of the time series $\{B_S(t_i; \mathbf{x}, c)\}_{i=1}^{n_t}$. A realisation of the maximum individual wave response $R_I|(\mathbf{X}=\mathbf{x}, C=c)$ is then obtained by applying \eqref{equation:per_wave_max}, using the set $t \in \{t_i\}_{i=1}^{n_t} \cap \mathcal{T}_0$ as an approximation to $\mathcal{T}_0$. Given conditioning crests $\{c_i\}_{i=1}^k$, the above procedure can be used to map $c_i \mapsto r_i$, for $i=1,\ldots, k$, obtaining a set of maximum responses $\{r_i\}_{i=1}^k$, for a given sea state $\mathbf{x}$.

\subsubsection{Importance sampling of simulated maximum responses} \label{section:importance_samp}

The procedure discussed in Section~\ref{section:conditional_wave_simulation} is used to obtain realisations $\{r_i\}_{i=1}^k$ of ${R_I|(\mathbf{X}, C)}$, for a set of $k$ conditioning crests $\{c_i\}_{i=1}^k$ and specified values of storm peak variables $\mathbf{X}$. These are then used in integral \eqref{equation:importance_sampling_target} to estimate the distribution $F_{R_I|\mathbf{X}}$ of the maximum response to an individual wave in sea state $\mathbf{X}$, assuming a Rayleigh distribution \eqref{equation:rayleigh_density_0} for $C|\mathbf{X}$. However, evaluation of integral \eqref{equation:importance_sampling_target} via Monte Carlo methods sampling from the Rayleigh density is inefficient in targeting the tail of the response distribution $F_{R_I|\mathbf{X}}$. Given our interest in the extreme structural response on the structure, we therefore employ the importance sampling approach described by \cite{towe2021efficient}, writing integral \eqref{equation:importance_sampling_target} as
\begin{equation} \label{equation:importance_sampling_definition}
    F_{R_I|\mathbf{X}} (r|\mathbf{x}) = \int_{r} F_{R_I|\mathbf{X}, C}(r|\mathbf{x}, c)   \frac{f_{C|\mathbf{X}}(c|\mathbf{x})}{g^{(\epsilon)}_{C|\mathbf{X}}(c|\mathbf{x})} g^{(\epsilon)}_{C|\mathbf{X}}(c|\mathbf{x}) \wrt{c},
\end{equation}
for $r  >0$, where $g^{(\epsilon)}_{C|(\mathbf{X}= \mathbf{x})}$ is the density of the $\text{Uniform}[0,\epsilon h(\mathbf{x})]$ distribution, for significant wave height $h(\mathbf{x})$ and some $\epsilon>0$. For a fixed number of conditional wave simulations, sampling of the conditioning crest $c$ from $g^{(\epsilon)}_{C|\mathbf{X}}$ ensures greater coverage of the feasible range of large crest elevations and of the induced maximum response than is achieved when sampling $c$ from $f_{C|\mathbf{X}}$. Therefore, the upper tail of the distribution $F_{R_I|\mathbf{X}}$ is more efficiently estimated using the sampling distribution $g^{(\epsilon)}_{C|\mathbf{X}}$. The value of $\epsilon$ in \eqref{equation:importance_sampling_definition} is selected so that $g^{(\epsilon)}_{C|\mathbf{X}}$ provides adequate coverage of the domain of $f_{C|\mathbf{X}}$, i.e., the exceedance probability
\begin{equation} \label{equation:exceedance_prob}
    \Pr\{C>\epsilon h(\mathbf{x})|\mathbf{X}=\mathbf{x}\} = \int_{c>\epsilon h(\mathbf{x})}f_{C|\mathbf{X}}(c|\mathbf{x})\wrt{c} = \exp\left( -8\epsilon^2\right),
\end{equation}
is sufficiently close to zero. We set $\epsilon=2$, which gives a value of probability \eqref{equation:exceedance_prob} in the order of $10^{-14}$. 

Integral \eqref{equation:importance_sampling_definition} is then estimated as follows. The $k$ conditional crests $\{c_i\}_{i=1}^k$ are sampled from the uniform proposal density $g^{(\epsilon)}_{C|\mathbf{X}}$. Corresponding realisations  $\{r_i\}_{i=1}^k$ of single-wave maximum responses are obtained using the procedure described in Section~\ref{section:conditional_wave_simulation}.  The distribution $F_{R_I|\mathbf{X}}$ is then estimated as
\begin{equation} \label{equation:importance_sampling}
\hat{F}_{R_I|\mathbf{X}} (r|\mathbf{x}) = \frac{\sum_{i=1}^k\mathbbm{1}_{\{r_i\leq r\}}\iota^{(\epsilon)} (c_i|\mathbf{x})}{\sum^k_{i=1} \iota^{(\epsilon)} (c_i|\mathbf{x})}, 
\end{equation}
for $r >0$, where $\iota^{(\epsilon)}(c|\mathbf{x}) = f_{C|\mathbf{X}}(c|\mathbf{x})/{g^{(\epsilon)}_{C|\mathbf{X}}(c|\mathbf{x})}$ is the importance sampling ratio and $\mathbbm{1}_{\{r_i\leq r\}}=1$ if $r_i\leq r$, zero otherwise, for $i=1,\ldots, k$. Estimate \eqref{equation:importance_sampling} is an empirical cumulative distribution function of the simulated responses, weighted to remove bias introduced from sampling crests from $g^{(\epsilon)}_{C|\mathbf{X}}$ rather than $f_{C|\mathbf{X}}$. We use estimate \eqref{equation:importance_sampling} to evaluate the distribution of maximum response per $L$-hour sea state using relation \eqref{equation:powering_up_integral}. Given an estimate for $f_{\mathbf{X}}$ obtained as in Section~\ref{section:conditional_extremes}, we may then calculate the marginal maximum response distribution using integral \eqref{eqn:forwardModelSimple}.

\section{Environmental contours} \label{section:contours}

\subsection{Overview of environmental contours}

Environmental contours provide a method of determining extremal conditions which are in some way related to an extreme structural response. These contours often make assumptions about the interaction between environment and response, usually regarding the shape of some failure boundary in the environment space such that environmental conditions beyond the boundary will result in structural failure. For instance, IFORM (e.g., \citealt{winterstein1993environmental}) contours assume a convex form for this boundary, whereas ISORM (e.g., \citealt{Chai2018}) assumes it to be concave. These assumptions may or may not be valid depending on the specific features of the structure type in question. Here, we outline the methodology of the IFORM contour (Section~\ref{section:IFORM}) and the fitting approach we employ to estimate it for our example dataset (Section~\ref{section:hierarchical}). 

\subsection{IFORM design contours} \label{section:IFORM}

Section~\ref{section:probability_failure} details how, given an estimate for the distribution of $R_L|\mathbf{X}$ from Section~\ref{section:methodology}, in principle we can evaluate the probability of structural failure $\Prb{\mathcal{F}_L|\mathbf{X}=\mathbf{x}}$ for a sea state (of duration $L$) with variables $\mathbf{X}=\mathbf{x}$. IFORM offers an approach to structural design which avoids direct calculation of $\Prb{\mathcal{F}_L|\mathbf{X}=\mathbf{x}}$, by attempting to make conservative assumptions. For an ocean environment represented by a set of random variables $\mathbf{X}$ transformed to independent standard Gaussian random variables $\mathbf{U}$, IFORM assumes that $\Prb{\mathcal{F}_L|\mathbf{U}=\mathbf{u}}$ is deterministic for all $\mathbf{u}$, taking values $\{0,1\}$, contrary to the failure probability discussed in Section~\ref{section:probability_failure} which takes any value in $[0,1]$. Writing the region of environmental space corresponding to failure $\mathcal{F}_L$ as $\mathcal{F}$, IFORM assumes that the boundary $\partial\mathcal{F}$ of $\mathcal{F}$ {is linear}, and lies tangential to a contour of constant transformed environmental density $f_\mathbf{U}$, making the assumed location of $\partial\mathcal{F}$ dependent on the joint Gaussian distribution $f_\mathbf{U}$. The assumption of a failure boundary of this type is typically conservative, in that estimates for $\Pr(\mathbf{U} \in \mathcal{F})$ using it have positive bias. 

The transformation of $\mathbf{X}\mapsto \mathbf{U}$ is achieved via the method of \cite{rosenblatt1952remarks}, which proceeds as follows. For storm peak variables $\mathbf{X}=(X_1, \ldots, X_p)$, suppose we can estimate the nested conditional distributions $F_{X_1}, F_{X_2|X_1}, F_{X_3|(X_1, X_2)}, \ldots, F_{X_p|(X_1, \ldots, X_{p-1})}$. Estimation of these distributions is non-trivial as it involves estimating a sequence of conditional dependence models with results dependent on the sequencing of the $p$ environmental variables; see Section~\ref{section:hierarchical} for an example approach. The Rosenblatt transformation maps a realisation $\mathbf{x}= (x_1, \ldots, x_p)$ of $\mathbf{X}$ to the realisation $\mathbf{u}=(u_1, \ldots, u_p)$ of $\mathbf{U}$ via $u_1 = \Phi^{-1} \{F_{X_1}(x_1)\}$ and $ u_j = \Phi^{-1}\left\{F_{X_j|(X_1, \ldots, X_{j-1})}(x_j|x_1, \ldots, x_{j-1})\right\}$, for $j=2, \ldots, p$, where $\Phi$ is the standard Gaussian cumulative distribution function.

In $\mathbf{U}$-space, contours of constant probability joint density correspond to boundaries of hyperspheres centred at the origin. In particular, the $P$-year IFORM contour in $\mathbf{U}$-space is the boundary of a hypersphere with radius 
\begin{equation}\label{equation:failure_index}
    \beta_I = \Phi^{-1}\left(1-\frac{1}{N_\text{an}P}\right),
\end{equation}
where $N_\text{an}$ is the average number of independent storm peak observations per annum. That is, the probability of a point lying outside the set enclosed by the $P$-year IFORM contour is $1/(N_\text{an} P)$. 
Consider failure region $\mathcal{F}_{\beta_I}$ with boundary $\delta \mathcal{F}_{\beta_I}$ tangential to the hypersphere centred at the origin with radius $\beta_I$. For any angle $\boldsymbol{\psi}$, in spherical polar coordinates for a point on the hypersphere, we have
\begin{equation}
    \Prb{\mathbf{U} \in \mathcal{F}_{\beta_I}} = \Prb{\sum_{i=1}^p w_i (\boldsymbol{\psi}) U_i > \beta_I},
\end{equation}
where $\sum^p_{i=1}w_i^2(\boldsymbol{\psi})=1$, due to the linearity of $\delta \mathcal{F}_{\beta_I}$ and it being tangential to the hypersphere at radius $\beta_I$. So, as $W=\sum_{i=1}^pw_i(\boldsymbol{\psi)}U_i \sim N(0,1)$ given the independence of $(U_1, \ldots, U_p)$, it follows that $\Prb{\mathbf{U}\in \mathcal{F}_{\beta_I}} = \Prb{W > \beta_I}$, which directly gives expression \eqref{equation:failure_index} for $\beta_I$.

 Once the environmental contour has been estimated in $\mathbf{U}$-space, it can be represented \linebreak in the original $\mathbf{X}$-space via the inverse Rosenblatt transformation $x_1 = F^{-1}_{X_1}\{\Phi(u_1)\}$ and  $x_j = F^{-1}_{X_j|(X_1, \ldots, X_{j-1})}\{\Phi (u_j)|x_1, \ldots, x_{j-1}\}$, for $j=2, \ldots, p$. {Unlike the hypersphere-shaped contours in the $\mathbf{U}$-space, these contours are not guaranteed to be convex (see Section~\ref{section:env_model}).} 
The procedure for construction of a $P$-year IFORM contour in terms of environmental variable $\mathbf{X}=(X_1, \ldots, X_p) $ is summarised in Algorithm~\ref{alg:form}.

\begin{algorithm}[H]
\caption{IFORM contour calculation for $\mathbf{X}=(X_1, \ldots, X_p)$} \label{alg:form}
 \hspace*{\algorithmicindent} \textbf{Input} Return period $P$; Average number of independent storm peaks per annum $N_\text{an}$; Estimates of distributions $F_{X_1}, F_{X_2|X_1}, F_{X_3|(X_1, X_2)}, \ldots, F_{X_p|(X_1, \ldots, X_{p-1})}$.\\
 \hspace*{\algorithmicindent} \textbf{Output} $P$-year IFORM contour.
\begin{algorithmic}[1]
\State Define $\beta_I = \Phi^{-1}\left(1-\frac{1}{N_\text{an}P}\right)$ where $\Phi$ is the standard normal cdf.
\State Obtain a set of $k$ equally spaced points $\{(u_{1(j)}, \ldots, u_{p(j)})\}^k_{j=1}$ on the hypersphere given by $u_{1}^2 + \ldots + u_{p}^2 = \beta_I^2$, using a regular grid of values of $\boldsymbol{\psi}$ in spherical polar coordinates.
\For{$j$ in $1, \ldots, k$}
\State Compute $x_{1(j)} = F_{X_1}^{-1}\{\Phi(u_{1(j)})\}$.
\State Compute $x_{2(j)} = F_{X_2|X_1}^{-1}\{\Phi(u_{2(j)})|x_{1(j)}\}$.
\State \vdots
\State Compute $x_{p(j)} = F^{-1}_{X_p|(X_1, \ldots, X_{p-1})} \{\Phi(u_{p(j)})|(x_{1(j)}, \ldots, x_{p-1(j)})\}$.
\EndFor
\Return $\{(x_{1(j)}, \ldots, x_{
p(j)})\}_{j=1}^k$ points along the $p$-dimensional IFORM contour.
\end{algorithmic}
\end{algorithm}
\subsection{Joint parametric models for storm peak variables} \label{section:hierarchical}
Construction of the IFORM contour for environmental variables $\mathbf{X}= (X_1, X_2) = (H_S, S_2)$ via Algorithm~\ref{alg:form} requires estimates for the marginal distribution $F_{H_S}$ and conditional distribution $F_{S_2|H_S}$ (while $F_{S_2}$ and $F_{H_S|S_2}$ could equally be used, we use the former to reflect past approaches). {Mirroring the hierarchical approach of \cite{winterstein1993environmental}, we select the GPD tail model \eqref{equation:marginal_model} for marginal $H_S$, and evaluate a range of parametric forms for the distribution of $S_2|H_S$, selecting the most appropriate based on an assessment of predictive performance. We estimate the model $S_2|H_S$ as follows.

We allow for two sources of flexibility: (a) the conditional distributional form for $S_2|H_S$, and (b) the nature of the parametric form for how the parameters of the distribution vary as a function of $H_S$. The distributions we consider in modelling step (a) are the Lognormal($\mu_L$, $\sigma_L$), as in \cite{winterstein1993environmental}, Gamma($\alpha$, $\beta$), Weibull($\lambda$, $k$) and the Generalised Extreme Value distribution GEV($\mu_G$, $\sigma_G$, $\xi$); see SM. We also consider conditional distributions fitted to the transformed negative steepness $S^-_2 = \min(S_2) - S_2$, enabling the right-hand tail of the distribution to be fitted to small $S_2$; see SM for a summary of all model combinations. In step (b), we impose linear, quadratic and exponential forms for the $S_2$ distribution parameters as functions of $H_S$. To assess the performance of candidate models for $S_2|H_S$ and $S_2^-|H_S$, we use a cross validation approach in which we evaluate the predictive likelihood of the models, focusing on the performance in the tail region (for large $H_S$). This proceeds as follows.

The sample $\{\mathbf{x}_i\}_{i=1}^{N_\text{all}}$ (see Section~\ref{section:motivation}) is partitioned into a `body' where $H_S \leq v$ and a `tail' where $H_S > v$ for $v>0$, denoted $\{\mathbf{x}_i^{B}\}_{i=1}^{N_B}$ and $\{\mathbf{x}_i^{T}\}_{i=1}^{N_T}$ respectively, with $N_B + N_T = N_\text{all}$. The tail portion $\{\mathbf{x}_i^{T}\}_{i=1}^{N_T}$ is itself partitioned into $K$ subsets $\{\mathcal{S}_j\}_{j=1}^K$ each of sizes $\lfloor N_T/K\rfloor$ or $\lceil N_T/K\rceil$. A $K$-fold cross validation is performed using training set $\{\mathbf{x}_i\}_{i=1}^N \setminus \mathcal{S}_j$ and test set $\mathcal{S}_j$ at fold $j\in \{1, \ldots, K\}$. That is, we always utilise the entirety of the body of the data within the training set alongside all but a single fold of the tail. Excluding a subset of the tail from the training data in this way is appropriate for estimation of the conditional distribution for $S_2|H_S$, but leads to biased estimation of the marginal distribution of $H_S$, so it is important to note estimation of the marginal model \eqref{equation:marginal_model} (see Section~\ref{section:conditional_extremes:marginal}) is carried out using the entire dataset. The predictive likelihood is then only calculated on extreme data points, and so measures the fit of each model to the extremes of the data.

We repeat the above process for $K=5, 10$ over values of $v=0, 0.8, 0.9$ to determine the sensitivity of model performance to the choice of extreme threshold. Setting $v=0$ recovers a standard cross validation approach for assessing fit to all of data, which we also include to ensure the best performing models fit the body of the data well. In addition, we also evaluate the model fit using AIC. The AIC and cross validation scores are each standardised by dividing by the number of observations for which we evaluate the optimised likelihood (negated when considering AIC, which is standardised over both terms). This standardisation results in a set of loosely comparable scores for each model across different threshold choices, and these scores for each model are averaged over values of both $K$ and $v$ to obtain a single score, referred to as the aggregate score (AS); see SM for details. The model with the largest AS is deemed to be the best predictive model for the data.

\section{Results} \label{section:results}

\subsection{Estimating the joint density of storm peaks} \label{section:results_density}

We employ the forward methodology of Sections~\ref{section:conditional_extremes} and \ref{section:response_dist} to estimate the environmental density $f_{\mathbf{X}}$ and response distribution $F_{R_L|\mathbf{X}}$ for our motivating dataset of $\mathbf{X}=(H_S, S_2)$ introduced in Section~\ref{section:motivation}. In turn, these are used to evaluate the distributions $F_{R_S}$ and $F_{R_A}$ (as in Sections~\ref{section:storm_max_defninition} and \ref{section:reduced}) and subsequently $\cdep$ (as in Section~\ref{section:condition_density_definition}). The $\cdep$ is then compared with various IFORM contour estimates using the methods of Section~\ref{section:contours}.
The purpose of the comparison of $\cdep$ and IFORM contours is to demonstrate that none of the hierarchical environmental model forms underpinning the IFORM contour provide adequate description of $\cdep$ for all example structures considered. That is, the IFORM contours estimated in general do not identify the correct region of environmental space responsible for extreme structural response. Therefore, evaluation of response along these IFORM contour boundaries will not provide reliable realisations of the desired $P$-year response. 

We model the joint environment $\mathbf{X}$ using the method of \cite{HffTwn04} discussed in  Section~\ref{section:conditional_extremes}, fitted to data in the region $A_\nu = \{\mathbf{X}\in \RR^2 : H_S > \nu\}$ for conditioning threshold $\nu$. We take $\nu = \tilde{F}_{H_S}^{-1}(0.95)$. Inspection of plots for the variability of conditional extremes model parameters with respect to threshold indicated this choice of threshold to be within the interval for which these parameters are invariant. Using this fitted model, the density $f_{\mathbf{X}}$ is estimated in $A_\nu$ as described in Section~\ref{section:conditional_extremes:density}. The density in the complement $\RR^2 \setminus A_\nu$ is modelled empirically. Since our interest lies in environments with large $H_S$ and associated structural responses, we are not concerned with (a) smooth estimation of this lower portion of the density and (b) the density in the region corresponding to large $S_2$ but small $H_S$.
\begin{figure}[h]
    \centering
    \includegraphics[width=0.6\textwidth]{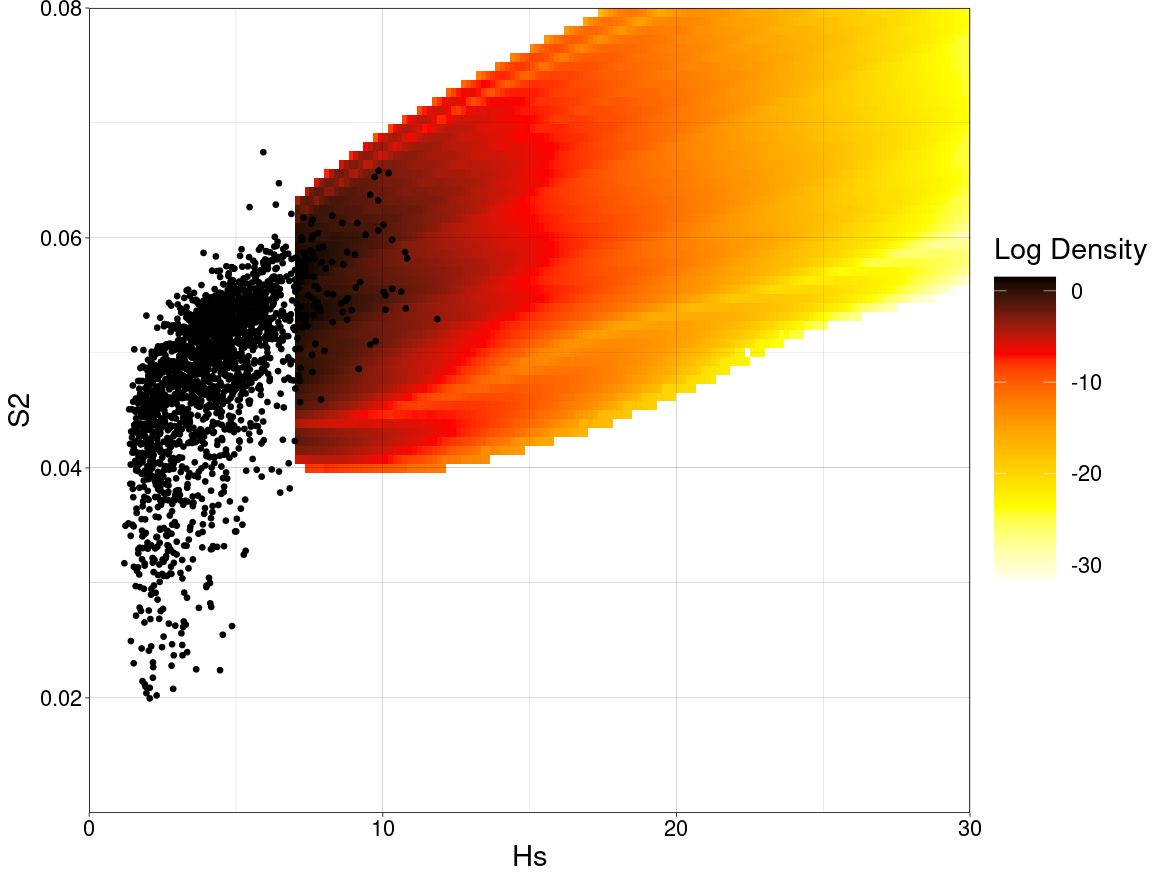}
    \caption{Storm peak density for $H_S$ [m] and $S_2$ in $A_\nu$ on log scale, estimated from the fitted conditional extremes model. Observations of storm peak $\mathbf{X}=(H_S, S_2)$ are shown as black dots.}
    \label{figure:storm_peak_dens}
\end{figure}

In the 2-dimensional case $\mathbf{X}=(H_S, S_2)$, the parameters of model \eqref{equation:ht_model} reduce to $\alpha \in [-1, 1]$ and $\beta \in (-\infty, 1]$ when conditioning on $H_S$. We find corresponding estimates $\hat{\alpha}=0.378$ and $\hat{\beta}=0.533$, as well as an estimate $\hat{\xi}=-0.02$ for the $H_S$ marginal shape parameter $\xi$ in \eqref{equation:marginal_model}, the latter indicating a near exponential upper tail for $H_S$. The effect of $\hat{\alpha}$ is seen in Figure~\ref{figure:storm_peak_dens}, as a positive trend in the resulting density estimate. This density also appears to agree with the shape of the data in the extreme region $\mathcal{A}_\nu$. 
{Figure \ref{figure:storm_peak_dens} also shows the presence of rays in the estimated density, caused by sampling from the kernel density of observed residual values (see \citealt{winter2017kth} or discussion in Section~\ref{section:conditional_extremes:joint}). The strength of these rays is determined by the value of the kernel smoothing parameter. Higher values provide stronger smoothing and therefore less prominent rays, however, too large a value will result in over-smoothing and thus less precise density estimates. When in doubt, we favour under-smoothing.}

\subsection{Selection of model form for the conditional distribution} \label{section:env_model}

For IFORM, we consider four distributional forms for each of $S_2|H_S$ and $S^-_2|H_S$ (summarised in Section~\ref{section:hierarchical}) each with two distribution parameters modelled as functions of $H_S$. We do not model the GEV shape $\xi$ as a function of $H_S$ as its estimation using maximum likelihood is difficult for finite samples; instead, we assume it to be an unknown constant. The variation of each of these eight distribution parameters with $H_S$ is represented by one of three parametric forms (linear, quadratic and exponential), giving a total of 72 combined candidate models for $S_2|H_S$ and $S^-_2|H_S$. These models are ranked using the AS, introduced in Section~\ref{section:hierarchical}, yielding results given in full in the SM. Table~\ref{table:avCVscores} summarises these results, showing the optimal model for each of $S_2|H_s$ and $S_2^-|H_S$, with unique $a, b, c \in \RR$ for each distributional parameter, constrained such that their respective domains are not violated. Standard errors are found as the sample standard deviation of the AS, evaluated over thirty replicates of each cross validation setting.
\begin{table}[ht]
\centering
\resizebox{\textwidth}{!}{
\begin{tabular}{|l|lll|l|}
  \hline
 Label
&\multicolumn{3}{|c|}{Distribution (with optimal functional form of parameters)}& \multicolumn{1}{c|}{AS}  \\
  \hline
   $\mathcal{C}_P^1$
&$S^-_2|\{H_S=h\} \sim \text{GEV}(\mu_G, \sigma_G, \xi)$ & $\mu_G =a + b \exp(ch)$ & $\sigma_G = a + b \exp(ch)$& \textbf{3.999 (0.002)}  \\ 
     $\mathcal{C}_P^2$
&$S_2|\{H_S=h\}\sim \text{Weibull}(k, \lambda)$ & $k = a + b c$&$\lambda = a(h+b)^2 + c$& \textbf{3.983 (0.003)}\\ 
       $\mathcal{C}_P^3$
&$ S^-_2|\{H_S=h\} \sim \text{Lognormal}(\mu_L, \sigma_L)$ & $\mu_L = a+ bh $&$\sigma_L =  a+ bh$& \textbf{3.963 (0.001)}\\ 
   $\mathcal{C}_P^4$
&$S^-_2|\{H_S=h\}\sim \text{Gamma}(\alpha, \beta)$ & $\alpha= a(h+b)^2 + c$&$\beta =  a+ bh$&  3.963 (0.006)  \\ 
   $\mathcal{C}_P^5$
&$  S^-_2|\{H_S=h\} \sim \text{Weibull}(k, \lambda)$ & $k = a + b \exp(ch)$&$\lambda = a + b \exp(ch)$& 3.897 (0.002)  \\ 
   $\mathcal{C}_P^6$
&$S_2|\{H_S=h\} \sim \text{Gamma}(\alpha, \beta)$ & $\alpha =  a+ bh$&$\beta =  a+ bh$&  3.873 (0.002)  \\ 
   $\mathcal{C}_P^7$
&$ S_2|\{H_S=h\} \sim \text{Lognormal}(\mu_L, \sigma_L)$ & $\mu_L = a(h+b)^2 + c$&$\sigma_L =  a + b \exp(ch)$& 3.824 (0.005)  \\ 
      $\mathcal{C}_P^8$&$S_2|\{H_S=h\} \sim \text{GEV}(\mu_G, \sigma_G, \xi)$ & $\mu_G = a + b \exp(ch)$&$\sigma_G =a(h+b)^2 + c$&  3.533 (0.000)  \\ 
   \hline
\end{tabular}}
\caption{AS (with standard error) for the optimal forms of each distribution considered for $S_2|H_S$ and $S^-_2|H_S$. Large values of AS indicate good performance. The three best performing models' scores are indicated in bold.  Complete results are given in the SM.}
\label{table:avCVscores}
\end{table}

The models in Table~\ref{table:avCVscores} are used with Algorithm~\ref{alg:form} to construct the IFORM contours in Figure~\ref{fig:all_contours}. Each contour corresponds to a return period of $P=1000$ years.  Contours are labelled $\mathcal{C}_P^1$ to $\mathcal{C}_P^8$ and ordered according to their AS, with the best fitting models having the lowest labelling.
All of the contour estimates provide plausible descriptions of the shape of the sample, but from an engineering design perspective, we note clear differences in the shape and position of the contours for larger $H_S$.
Even so, the three highest ranking models generate contours which agree to a reasonable degree in all regions. These three contours also appear visually to be the best descriptions of shape of the data. In comparison, the other contours do not agree in the region of large $H_S$, and fail to capture the shape of the main body of the data. We therefore select the highest three ranking contours as the best representations of IFORM to compare to $\cdep$ in Section~\ref{section:results_CDE}. 
\begin{figure}[h]
    \centering
    \includegraphics[width=0.65\textwidth]{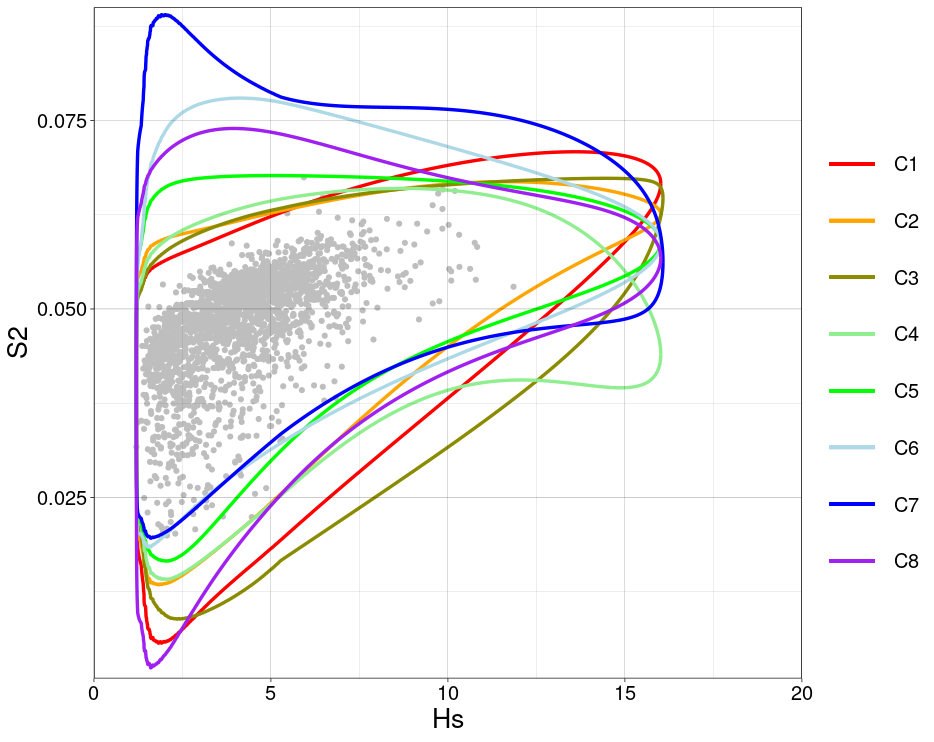}
    \caption{IFORM contours in $H_S$ [m] and $S_2$ constructed from the models in Table~\ref{table:avCVscores}, corresponding to an exceedence probability of $p=10^{-3}/73$, or a return period of $P=1000$ years for data with $N_\text{an}=73$ observations per annum. Contours $\mathcal{C}_P^1$ to $\mathcal{C}_P^8$ are listed and coloured in order of decreasing performance AS, from red to purple, with contour $\mathcal{C}^i_P$ labelled C$i$ for $i=1\ldots, 8$.}
    \label{fig:all_contours}
\end{figure}
\subsection{Estimating the conditional density of associated environmental variables} \label{section:results_CDE}

\subsubsection{Estimation of $\cdep$ for example structure models}

We evaluate $\cdep$ for three examples of the stick structure model (Section~\ref{section:conditional_wave_simulation}), denoted A, B \& C. These structures assume different values for drag and inertia coefficients $c_d(z)$ and $c_m(z)$ along their height $z$, as shown in Table~\ref{table:structures}. Structure A represents the simplest stick structure, with homogeneous drag and inertia coefficients along its entire height, i.e., $c_d(z) = c_m(z) = 1$ for all $z\in [-100, 150]$. We mimic wave-in-deck loads for structure B, with $c_d(z)=c_m(z)=100$ increased for a portion of the structure above sea level. A portion of structure C near the sea bed incurs increased load.
\begin{table}[h]
    \centering
\begin{tabular}{|l|lllc|}
\hline
                           &                        & \multicolumn{3}{c|}{Value of $c_d(z)=c_m(z)$}                                                                       \\ \hline
                           & \multicolumn{1}{l|}{}  & \multicolumn{1}{l|}{$5 < z \leq 15$} & \multicolumn{1}{l|}{$-95 < z \leq -85$} & Elsewhere \\ \cline{2-5} 
\multirow{2}{*}{Structure} & \multicolumn{1}{l|}{A} & \multicolumn{1}{c|}{1}                     & \multicolumn{1}{c|}{1}                   & 1         \\
                           & \multicolumn{1}{l|}{B} & \multicolumn{1}{c|}{100}                   & \multicolumn{1}{c|}{1}                   & 1         \\
                           & \multicolumn{1}{l|}{C} & \multicolumn{1}{c|}{1}                     & \multicolumn{1}{c|}{100}                 & 1         \\ \hline
\end{tabular}
    \caption{The drag and inertia coefficients of \eqref{equation:morison_load} with  $c_d(z) = c_m(z)$ for all $z$, for varying $z$ associated with stick structures A, B \& C.}
    \label{table:structures}
\end{table}

Figure \ref{fig:just_cond_dens} shows the corresponding estimates for $\cdep$ for $P=1000$ years. The shape and position of the conditional density varies between structures, due to their differing loading characteristics. For structure C in particular, the conditional density extends to larger $H_S$, and over a wider interval of $S_2$; we comment further on this feature in the discussion of environmental contours in Figure~\ref{fig:fail_probs_trio}. \red{That is, Figure 4 demonstrates that more than one region of the ($H_S$, $S_2$) domain contributes to the the distribution of $P$-year response, particularly for structure C.}

\begin{figure}[h]
    \centering
    \includegraphics[width=\linewidth]{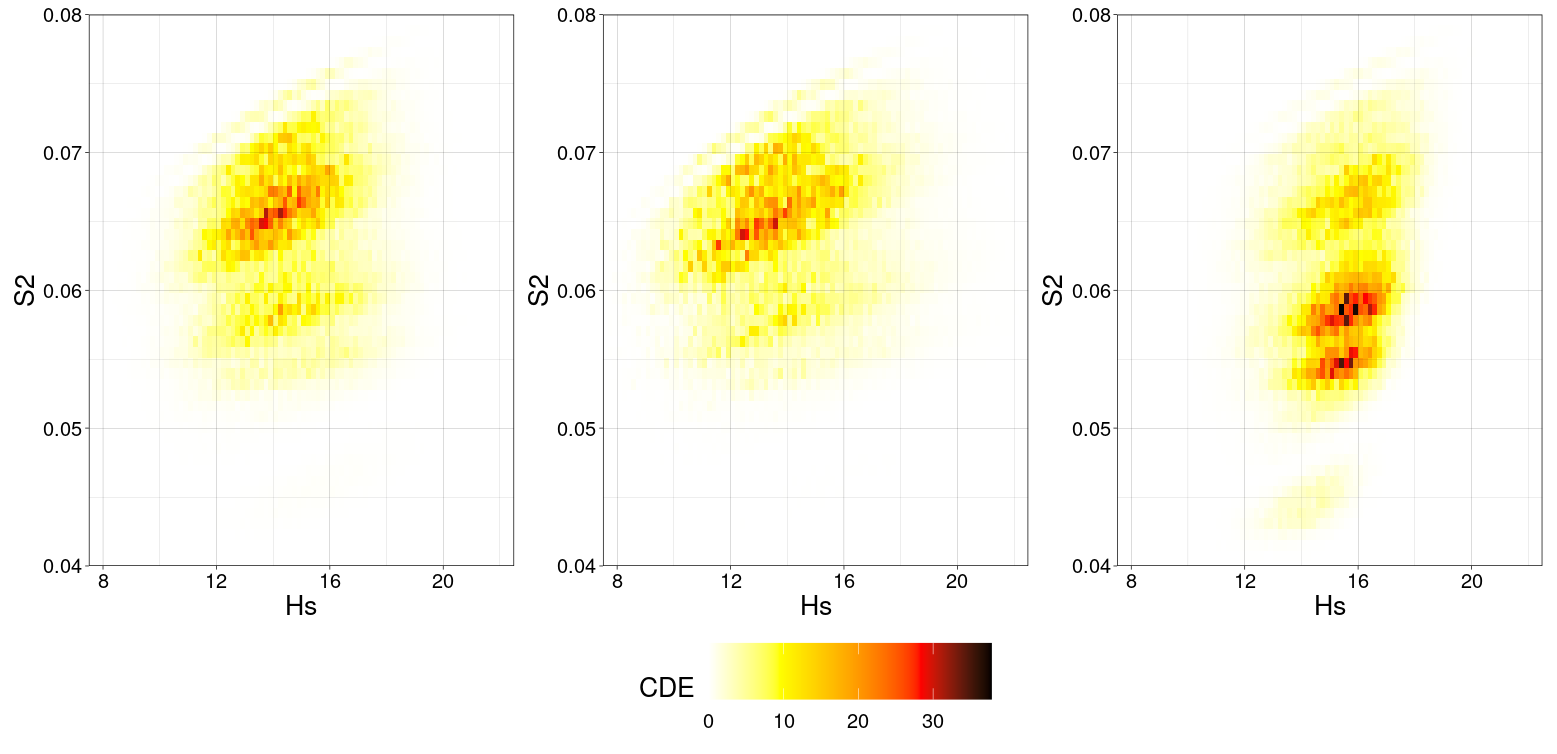}
    \caption{Estimated $\cdep$ for the three structure models given in Table~\ref{table:structures} (left to right A, B \& C), for a period of $P=1000$ years. These are evaluated for the example dataset given in Section~\ref{section:motivation}, using the methods of Section~\ref{section:methodology}.}
    \label{fig:just_cond_dens}
\end{figure}

\subsubsection{Comparison between $\cdep$ and IFORM contours}

Provided we have accurate models for the series of $p$ nested conditional distributions \linebreak  $F_{X_1}, \ldots, F_{X_p|(X_1, \ldots, X_{p-1})}$, and provided that the assumptions underlying IFORM are valid, the $P$-year IFORM contour gives design points at which evaluation of a response model will provide conservative estimates of the $P$-year response, as indicated by \cite{winterstein1993environmental}. That is, it aims to provide environmental conditions at least as extreme as those which induce the $P$-year response for specific structures. It is natural therefore to consider assessing IFORM contour performance using $\cdep$, since the latter provides asymptotically-justified estimates of the environmental conditions corresponding to the $P$-year response, obtained from application from the forward approach of Section~\ref{section:methodology}. 

Assuming that $\cdep$ provides a valid estimate of the environment density conditional on the $P$-year response, we reason that a well-estimated IFORM contour $\mathcal{C}_P$ should intersect with $\cdep$, and that the discrepancy between $\mathcal{C}_P$ and $\cdep$ points to inadequacy of the IFORM methodology, potentially due to a mis-specified parametric environmental model or invalid IFORM assumptions regarding the characteristics of the wave-structure interaction for the application at hand. For instance, see Figure~\ref{fig:cond_dens_trio}, which shows the estimated $\cdep$ for structures A, B and C, for $P=1000$. Contour $\mathcal{C}_P^2$ in the left panel overlaps with darker points of $\cdep$ so provides estimates of the $P$-year response which agree with those obtained from the forward approach, meaning it performs well in this region of the contour. The centre panel of Figure~\ref{fig:cond_dens_trio} shows $\mathcal{C}_P^1$ taking $(H_S, S_2)$ values more extreme than the darker points of $\cdep$, indicating that the contour will provide estimates for the $P$-year response more extreme than those obtained from the forward approach, i.e., the conservative outcome intended. Conversely, the right panel shows $\mathcal{C}_P^2$ taking $(H_S, S_2)$ values less extreme than the darker points of $\cdep$, suggesting estimates for the $P$-year response obtained from points along this contour will be smaller than those obtained from the forward approach and hence they fail to give conservative estimates, which is contrary to their claimed properties. More generally, the majority of the presented IFORM contours exhibit over-conservatism in regions of small $S_2$, i.e., they lie in regions of the environment space with more severe $H_S$ than in the regions with non-zero $\cdep$, and this could lead to substantial over-design if this region of the environmental space was important. 

We observe from the above analysis of Figure~\ref{fig:cond_dens_trio} that the overlap between the region of non-zero $\cdep$ and the region $\mathcal{A}_P$ bounded below by the IFORM contour $\mathcal{C}_P$ relates to the level of conservatism, and therefore is a good scalar metric for measuring performance of IFORM contours relative to the more strongly justified forward approach. If this overlap is zero, then the contour lies in a region less extreme than the region of non-zero $\cdep$, indicating non-conservatism. Conversely, if the overlap includes all of the non-zero region of $\cdep$, the contour appears to be conservative. To formalise this observation, we define the following metric, which quantifies the level of this overlap. Consider
\begin{equation}\label{equation:overlap_integral}
    \zeta(P, \mathcal{A}_P) = 2\int_\mathbf{x} \mathbbm{1}_{\mathcal{A}_P}(\mathbf{x}) f_{\mathbf{X}|R_L}(\mathbf{x}|r_P) \wrt{\mathbf{x}} - 1,
\end{equation}
for $f_{\mathbf{X}|R_L}$ as in Section~\ref{section:condition_density_definition}, with $\mathbbm{1}_{\mathcal{A}_P}(\mathbf{x})=1$ if $\mathbf{x} \in \mathcal{A}_{P}$ and $\mathbbm{1}_{\mathcal{A}_P}(\mathbf{x})=0$ otherwise, and where $r_P$ is the $P$-year response of the structure. The metric $\zeta(P, \mathcal{C}_P)$ takes values in [-1, 1]. Here, $\zeta(P, \mathcal{C}_P)>0$ indicates conservatism of $\mathcal{C}_P$ due to overestimation of the $P$-year response, $\zeta(P, \mathcal{C}_P)< 0$ indicates underestimation of the $P$-year response and so non-conservatism, and $\zeta(P, \mathcal{C}_P)\approx0$ indicates accurate estimates for the $P$-year response and so optimal overall structural design given by IFORM. 

To justify our findings regarding $\zeta(P, \mathcal{C}_P)$ in a general case, we apply the following heuristic argument. First, we observe from Figure~\ref{fig:just_cond_dens} that $\cdep$ exhibits approximate marginal symmetry with respect to both $H_S$ and $S_2$ (i.e., little marginal skew). We also assume $\mathcal{A}_{P_1} \subset \mathcal{A}_{{P_2}}$ for any $P_2>P_1\geq 1$, and that a contour $\mathcal{C}_{P}$ varies smoothly with $H_S$ and $S_2$ for a given $P$. Then, (a) if $\zeta(P, \mathcal{C}_P)$ $\approx$ 0, the integral over $\cdep$ within $\mathcal{A}_P$ is approximately equal to the integral over $\cdep$ within $\mathcal{A}^c_P$. We interpret this as indicating that points on $\mathcal{C}_P$ coincide with high values of $\cdep$, as in the cases seen in Figure~\ref{fig:cond_dens_trio}, hence structural responses corresponding to points on $\mathcal{C}_P$ will be of similar magnitudes to those from the high density regions of $\cdep$. However, (b) if $ \zeta(P, \mathcal{C}_P)>0$, $\mathcal{A}_P$ contains the environmental region where $\cdep$ is non-zero, hence points on $\mathcal{C}_P$ produce structural responses beyond the $P$-year level, resulting in conservative design using IFORM contour $\mathcal{C}_P$. Further, (c) if $ \zeta(P, \mathcal{C}_P) <0 $, the intersection between $\mathcal{A}_{P}$ and the non-zero $\cdep$ region of the environment space is negligible. Given our assumptions, this arises only when $\mathcal{A}_{P}$ occupies a region of the environmental space less extreme that that with non-zero $\cdep$. Under these circumstances, points on $\mathcal{C}_P$ will produce structural responses corresponding to return periods less that $P$, resulting in a lack of conservatism in design.

Estimates for $\zeta(P,\mathcal{C}_P)$ in Table~\ref{table:overlap_ints} support this interpretation, relative to our estimates in Figure~\ref{fig:cond_dens_trio}. For structure A, 
only contour $\mathcal{C}_P^2$ appears to pass through the highest density region of $\cdep$. The other contours lie beyond the highest density region of $\cdep$, corresponding to $\zeta(P,\mathcal{C}_P)>0$. Observations for structure B are similar, since $\cdep$ does not vary considerably between structures A and B. For structure C, relative to A and B, the highest density regions in $\cdep$ occur at higher $H_S$ but lower $S_2$. Hence, despite no change in the locations of contour $\mathcal{C}_P^j$ ($j=1,2,3$), now only contour $\mathcal{C}_P^3$ is conservative. Contours $\mathcal{C}_P^1$ and $\mathcal{C}_P^2$ are clearly non-conservative, as confirmed in Table~\ref{table:overlap_ints}.

\begin{table}[H]
\centering
\begin{tabular}{|c|ccc|}
\hline
\multirow{2}{*}{Contour} & \multicolumn{3}{c|}{$\zeta(P, \mathcal{C}_P)$}                              \\ \cline{2-4} 
                             & \multicolumn{1}{c|}{Structure A}         & \multicolumn{1}{c|}{Structure B} & Structure C \\ \hline
$\mathcal{C}_P^1$& \multicolumn{1}{c|}{0.353} & \multicolumn{1}{c|}{0.475}  &   -0.259\\
 $\mathcal{C}_P^2$& \multicolumn{1}{c|}{-0.125} & \multicolumn{1}{c|}{-0.016}  &   -0.539\\
$\mathcal{C}_P^3$& \multicolumn{1}{c|}{0.176} & \multicolumn{1}{c|}{0.200}  &  0.042\\
\hline
\end{tabular}
\caption{Estimates of $\zeta(P, \mathcal{C}_P)$ from \eqref{equation:overlap_integral} for IFORM contours corresponding to the three best fitting models for $H_S$, $S_2$ from Table~\ref{table:avCVscores}, for each of structures A, B and C from Table~\ref{table:structures}.}
\label{table:overlap_ints}
\end{table}

\begin{figure}[h]
    \centering
    \includegraphics[width=\linewidth]{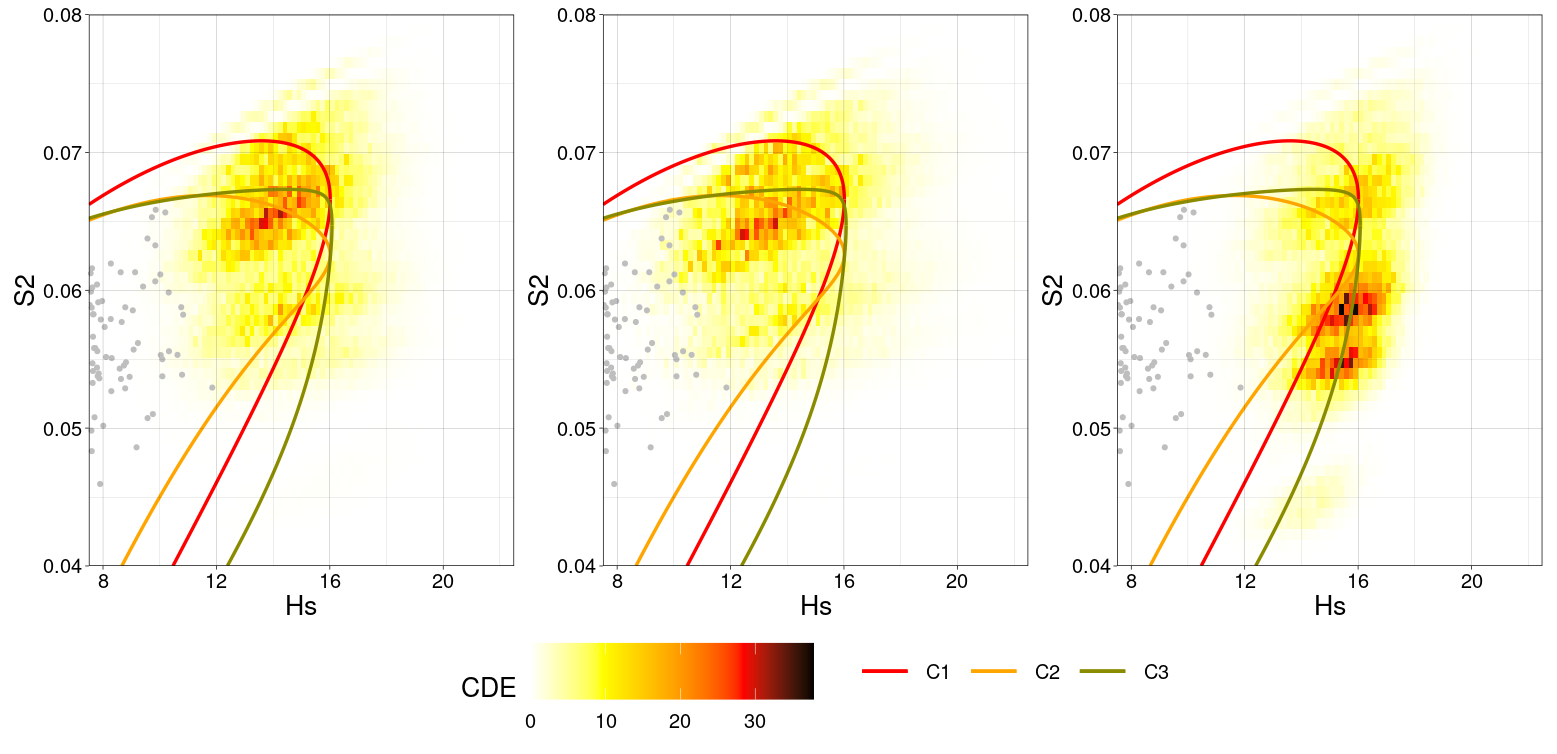}
    \caption{Density of storm peak sea state parameters $(H_S, S_2)$ conditioned on observing the 1000-year marginal response on stick-type structures. Overlaid are the three highest scoring IFORM fits by AS, with contour $\mathcal{C}^i_P$ labelled C$i$ for $i=1\ldots, 3$. The colouring of the contours (red, orange, grey) indicates the order of ranking, in terms of decreasing predictive performance.}
    \label{fig:cond_dens_trio}
\end{figure}

\begin{figure}[h]
    \centering
    \includegraphics[width=\linewidth]{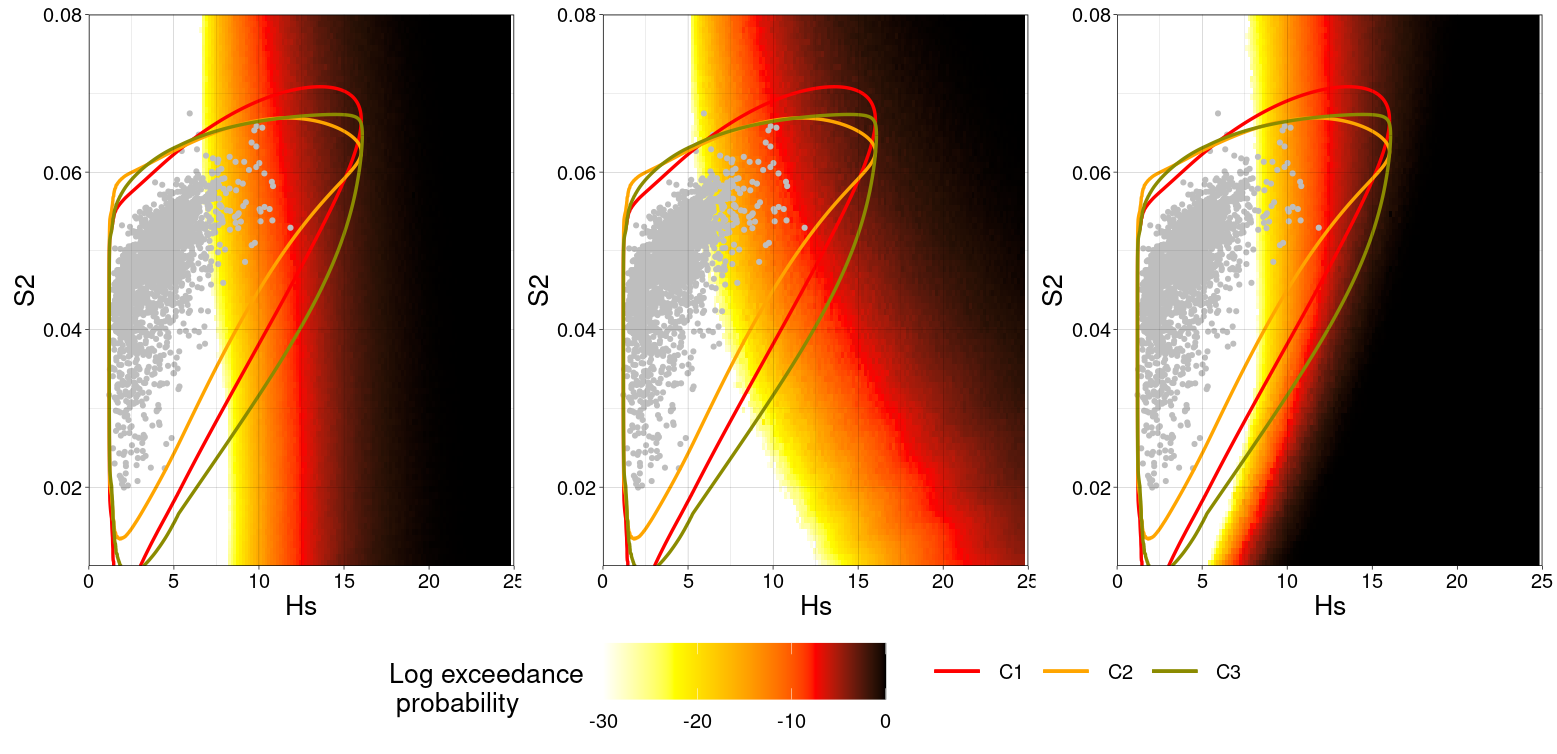}
    \caption{Estimated log probability of exceeding the $P$-year response for $P=1000$ within a 3-hour sea state as a function of $H_S$ and $S_2$, for structures A, B and C (left to right), obtained using the foward method of Section~\ref{section:methodology}. Overlaid are IFORM contours for the three best-fitting environmental models, with contour $\mathcal{C}^i_P$ labelled C$i$ for $i=1\ldots, 3$. The original sample of storm peak $(H_S, S_2)$ is shown as grey dots. }
    \label{fig:fail_probs_trio}
\end{figure}

Figure~\ref{fig:fail_probs_trio} shows the estimated log probability of exceeding the $P$-year response 3-hour sea state, given by $\log \{1-F_{R_L|\mathbf{X}}(r_p|\mathbf{x}) \}$ where $F_{R_L|\mathbf{X}}$ is estimated as in Section~\ref{section:methodology}, for $P=1000$ and for all $\mathbf{X}=(H_S, S_2)$ within a subset $[0, 25] \times [0, 0.08]$ of the environment space $\RR^2$. IFORM contours using the highest ranking environmental model are overlaid. The panels show that the probability of exceeding $P$-year response for given environmental conditions varies with structure type, resulting in differing locations of a `frontier' where this probability becomes non-zero (i.e., where the log probability exceeds roughly -30), corresponding to the shaded regions on the plots. For structure A, the frontier lies along constant {$H_S$}, indicating that $H_S$ alone effects the base shear induced on the structure. For structures B and C, we see convex and concave curvature of the frontier respectively, indicating that these structures are more susceptible to high- and low-steepness conditions respectively. These differences in the underlying wave-structure interaction are reflected in the positions of $\cdep$ in Figures \ref{fig:just_cond_dens} and \ref{fig:cond_dens_trio}, but not in the locations of the IFORM contours which remain the same across all structures. This change in frontier location is therefore the cause of different contour performances across structures, seen in Table~$\ref{table:overlap_ints}$ and Figure~\ref{fig:cond_dens_trio}. For instance, contours $\mathcal{C}_P^1$ and $\mathcal{C}_P^2$ give non-conservative estimates for the $P$-year response on structure C because they fail to account for the concave curvature of the frontier see in the third panel of Figure~\ref{fig:fail_probs_trio}, and so not react to the increased severity of response seen at lower values of $S_2$.
{In summary, none of the three contours $\mathcal{C}_P^1$, $\mathcal{C}_P^2$, $\mathcal{C}_P^3$ provides the same level of conservatism for all three example structures. Indeed some contours, although fitting the sample data well, are non-conservative for design.}

\section{Discussion} \label{section:discussion}

{Offshore structures are subjected to extremes of environmental conditions (in the current context, significant wave height $H_S$ and wave steepness $S_2$), making structural risk assessment a critical step in the design process. Ideal methods combine models for environmental extremes with direct estimation of the environment-structure interaction from fluid loading. We demonstrate an efficient approach to estimating the distribution of extreme base shear on three stick structures, in a central North Sea environment. Environmental modelling is carried out using a combination of asymptotically justified models for univariate \citep{davison1990models} and conditional extremes \citep{HffTwn04}. Simulation of short term conditional wave fields then exploits the conditional method of \cite{TylEA97} with efficient sampling \citep{TowEA23} to estimate the distribution of Morison base shear (see \citealt{MrsEA50}). 

Using the full probabilistic analysis, we estimate the conditional density of the environment variables ($\cdep$) given occurrence of the $P$-year response. We adopt $\cdep$ as a diagnostic tool with which to evaluate the usefulness of any approach which claims to identify regions of the environmental space associated with extreme structural response.

Due to the perceived computational complexity of the forward approach, metocean engineers often use environmental contour methods as computationally simpler alternatives to approximate a full probabilistic analysis. These characterise the joint environment only, and then rely on assumptions about the nature of the environment-structure interaction to be useful for design (see \citealt{Ross2020, HslEA21} for a recent review). The usual argument in favour of adopting an environmental contour approach is that the contour can be estimated without knowledge of the structure; this is correct. However, once the contour is applied to assess the reliability of an actual structure, an assumption is made that the contour boundary is informative for environmental conditions likely to generate extreme structural responses. {In this work, we demonstrate that this assumption is not correct in general.} 

Results comparing $\cdep$ with IFORM contour estimates highlight two deficiencies of the IFORM method, and hence of design contours in general. First, although design contours are intended to be conservative by construction, they are unable to reflect the type of structure under consideration and so may perform well for one structure but not for another. Second, identification of a good environmental model underpinning the contour is challenging, and there is considerable variability in the location of the contour due to the choice of environmental model; current practice tends to ignore this source of uncertainty. More generally, IFORM represents just one of a number of possible variants of environmental design contours (see \citealt{Ross2020, HslEA21, mackay2021marginal, mackay2023model}); it is often not clear which contour is most appropriate in a given application. ($\cdep$ allows us to identify the most appropriate contour approach for a given application, but at the cost of undertaking the full probabilistic analysis; in this instance, we would always choose the full probabilistic analysis in favour of the contour approach.)}

\red{Within the forward model, we adopt the conditional extremes model of \cite{HffTwn04} to estimate the joint distribution of extremes only of environmental variables. Using the forward model, we estimate the conditional density of the environment ($\cdep$), given an extreme response corresponding to a given return period $P$. We then use $\cdep$ to compare with different environmental contours corresponding to the same return period. We use the IFORM procedure to estimate the environmental contour, which requires a statistical model for the full joint distribution of the environmental variables. We explore a range of hierarchical models for this purpose, reflecting the model forms typically used in the ocean engineering literature; these typically do not include the conditional extremes model. In particular, we establish that fitting of hierarchical models for $H_S$ and $S_2|H_S$ is a large source of uncertainty in  environmental contour estimation (see also e.g. \citealt{de2022quantitative}). We acknowledge that in future work it would also be interesting to explore the conditional extremes model for contour estimation further (following the work of e.g. \citealt{jonathan2010joint, jonathan2012joint, jonathan2014estimation, Ross2020} or \citealt{TowEA24}). In fact, \cite{jonathan2010joint} already provides a direct comparison of environmental contours estimated using the $H_S$, $T_P|H_S$ hierarchical model with that estimated using the conditional extremes model, where $T_P$ is the spectral peak period. See also \cite{tendijck2023extremal} for a discussion on extremal characteristics of hierarchical models.}

We employ techniques for selecting the extreme thresholds in the models of \cite{davison1990models} and \cite{HffTwn04} that can lead to subjective choices for each. Recent work by \cite{murphy2023automated} provides an automated approach to threshold selection that eliminates this subjectivity in the univariate case. Additional methods for handling the choice of conditioning threshold for the model of \cite{HffTwn04} exist, such as testing for the independence between exceedance and residual values (discussed by \citealt{jonathan2012joint}) or bootstrap sampling to quantify the uncertainty in parameter estimates due to threshold choice (see e.g., \citealt{jonathan2010joint}). Future analysis could therefore be improved by developing methods to automate the choice of conditional model threshold which incorporate the aforementioned techniques, to be used alongside the univariate method of \cite{murphy2023automated}.

In this work, simple stick structures are considered, with a response dependent on only two environmental variables ($H_S$ and $S_2$). We believe that this framework provides sufficiently realistic wave surface and kinematic models for our structure types. Despite this, future analysis might benefit from the inclusion of more complex structure models and wave-structure interactions. In reality, there are factors such as wind and current, alongside additional directional and seasonal covariate effects, not accommodated in this work. Structure models which include effects such as local loading and wave breaking could also be utilised, alongside improved models for the environment itself. For example, linear wave theory may be extended by transforming linear wave characteristics to their respective non-linear equivalents following the approach outlined in \cite{LOADS-Summary1} and \cite{LOADS-Summary2}. Use of more complex structure and environment models, however, incurs higher computational cost, and so an approach for efficient estimation of $\cdep$ that avoids the need for numerical simulation from fluid loading models may be desirable. Moreover, adoption of methods of full probabilistic structural design must be undertaken with care, to ensure rational evidence-based evolution of design procedures. For example, \cite{LOADS-NORSOC} identifies that a number of the features of the methodology of the LOADS joint industry project \citep{LOADS-Summary1, LOADS-Summary2} are either not compatible with the NORSOK standard, or not yet sufficiently well stress-tested for adoption within the standard. From this perspective, contours retain the advantages of being less computationally costly to employ, whilst also only requiring knowledge of the joint environment. 

\red{One approximate approach to reduce the bias in extreme response and associated risk statistics estimated from an environmental contour is to calibrate contour characteristics for a specific structural archetype. The necessary calibration would be estimated by applying the full forward probabilistic analysis for the structural archetype, then adjusting contour characteristics and/or associated calculations to reduce the observed bias in extreme response, or any other statistic of interest, estimated using the contour; indeed, the CDE might prove a useful basis for contour calibration. It might be possible to apply different calibration corrections to the contour in different parts of the environmental space in a systematic manner, with reduced need for user judgement, so that the contour better mimics CDE. The appropriate calibration could then be used with an environmental contour applied to that specific archetype. Note that this adjustment of contour characteristics to accommodate structural characteristics, is quite different to contour adjustment for short-term environmental variability (recommended by some standards; e.g., \citealt{standard2017actions}), and to adjustment of contour estimates made from serially-correlated data (e.g., \citealt{de2022quantitative}).}

 \red{During the review process, one referee queried (a) the usefulness of $\cdep$ as a diagnostic for environmental contours, and (b) whether the arguments given in the current work are an adequate reflection of the dangers (or otherwise) of environmental contour methods for design. We believe that the novelty of $\cdep$ as a diagnostic for environmental contours stems from its ability to capture the specific regions of the environment space responsible for extreme structural response, in a clear and systematic fashion with little need for user judgement. In contrast, current evaluation (and calibration) of environmental contours focuses only on comparison of responses calculated at user-selected points on the contour frontier with the relevant percentiles of the marginal short-term response distribution (e.g., \citealt{Ross2020}). The additional information provided by $\cdep$, in our opinion, provides improved qualitative understanding together with a basis for systematic quantification of the drawbacks of (IFORM) environmental contours when applied to different structural archetypes.} The current analysis illustrates the benefits of full probabilistic structural analysis relative to approximate analysis using environmental contours. Wherever possible, we recommend the application of full probabilistic structural design, \red{or alternatively of contour methods carefully calibrated for specific structural archetypes using full probabilistic analysis, potentially using the conditional density of the environment as a basis for calibration.}

\section*{Acknowledgements}
The work was completed while Matthew Speers was part of the EPSRC funded STOR-i centre for doctoral training (grant no. EP/S022252/1), with part-funding from the ARC TIDE Industrial Transformational Research Hub at the University of Western Australia. The authors wish to acknowledge the support of colleagues at Lancaster University and Shell.

\bibliographystyle{elsarticle-harv} 
\bibliography{bibli.bib}

\newpage 

\setcounter{section}{0}

\begin{center}
    \Large
    \textbf{Supplementary Material to `Estimating Metocean Environments Associated with Extreme Structural Response'}
\end{center}

\renewcommand{\theequation}{S.\arabic{equation}}
\renewcommand{\thesection}{S\arabic{section}} 
\renewcommand{\thefigure}{S\arabic{figure}}
\setcounter{figure}{0}
\renewcommand{\thetable}{S\arabic{table}}
\setcounter{table}{0}

\section*{Overview}

Here we present further details on aspects of the methodology covered in the main text. Section~\ref{section:univariate_threshold} describes the diagnostic techniques used to select the marginal exceedance thresholds for the generalised Pareto distribution (GPD) tail model (4) of Section~3.2.2. In Section~\ref{section:lin_wave_extra}, we provide further details on the origin of the wave kinematics equations given in Section 3.3.2, as well as information on the efficient simulation of wave fields from this model. Section~\ref{section:cv_scores_full} gives the specific parametric form of the models considered for $S_2|H_S$ and $S_2^-|H_S$ in Section~4.3, as well as the complete results summarised by Table~1 in Section~5.2.

\section{Univariate threshold selection for $H_S$ modelling}\label{section:univariate_threshold}

The suitable threshold for the modelling of $H_S$ and $S_2$ via (4) in Section~3.2.2 is selected using standard extreme value diagnostics, such as stability plots and mean-residual-life plots (see \citealt{coles2001introduction} for examples). These methods are summarised below. 

First, we consider the choice of threshold $u_{H_S}$ for $H_S$. Figure~\ref{fig:stability_plots} shows the values of the GPD standardised scale ${\sigma}^*=\sigma_u-\xi u$ (for $\sigma_u$ estimated with threshold $u>0$) and shape parameter $\xi$ for choices of threshold non-exceedance probability $p_{u_{H_S}} = \tilde{F}_{H_S}(u_{H_S})$, alongside their respective 95\% confidence intervals, obtained using block bootstrapping. The estimates of each parameter appear stable beyond the 0.8\textsuperscript{th} percentile. We thus select $u_{H_S}= \tilde{F}_{H_S}^{-1}(0.8)$ as the exceedance threshold for the marginal modelling of $H_S$. We select threshold $u_{S_2}$ for marginal modelling of $S_2$ using the same approach. Figure~\ref{fig:s2_plots} shows the equivalent plots obtained when fitting to $S_2$. Again, we see stability of parameter estimates for thresholds with exceedance probability past 0.8, and so we select $u_{S_2} = \tilde{F}_{S_2}^{-1}(0.8)$ as the exceedance threshold for the marginal modelling of $S_2$. 

We verify these threshold choices using mean residual life plots (see \citealt{coles2001introduction} for details). The form of the GPD implies that a linear trend in mean excesses with respect to threshold will occur above thresholds satisfying the model conditions. Figure~\ref{fig:mrl_plots} shows that there is a linear trend in mean excess point estimates when conditioning on thresholds above the 0.8\textsuperscript{th} percentile for both $H_S$ and $S_2$.
\begin{figure}[h]
\centering
\includegraphics[width=0.7\textwidth]{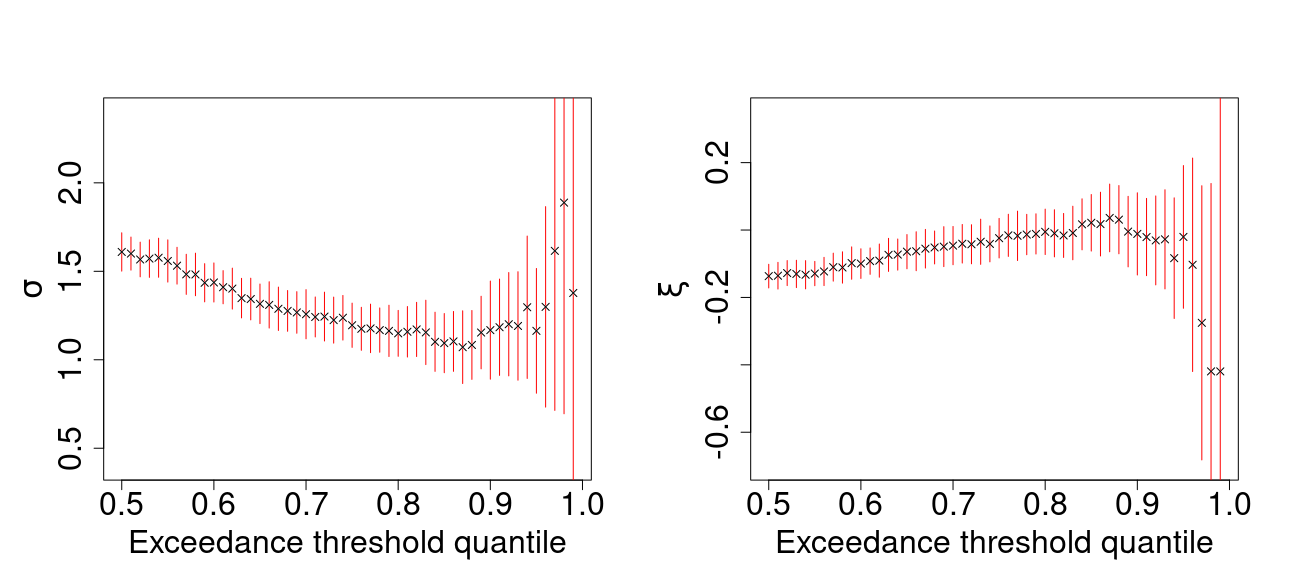}
\caption{Stability plots for estimated values of the GPD scale $\sigma$ and shape $\xi$ parameters when fitting to $H_S$ peak data, with respect to varying levels of the conditioning threshold percentile. Point estimates obtained via maximum likelihood estimation are marked in black. Corresponding 95\% confidence intervals are shown in red.}
\label{fig:stability_plots}
\end{figure}
\begin{figure}[h]
    \centering
    \includegraphics[width=0.7\textwidth]{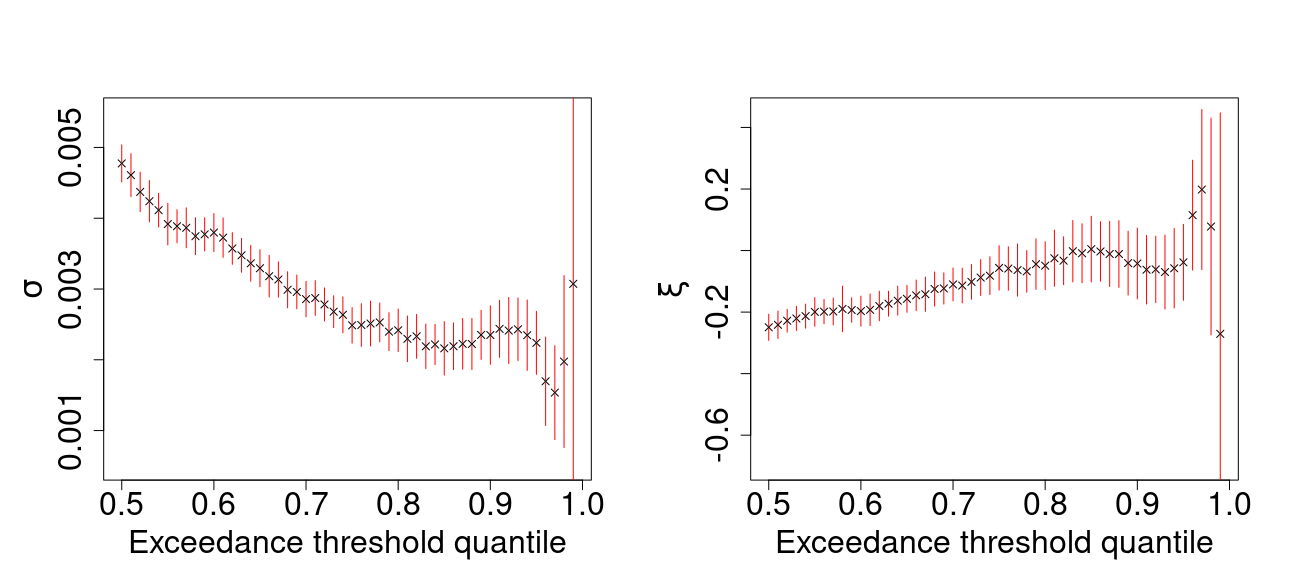}
    \caption{Stability plots for estimated values of the GPD scale $\sigma$ and shape $\xi$ parameters when fitting to $S_S$ peak data, with respect to varying levels of the conditioning threshold percentile. Point estimates obtained via maximum likelihood estimation are marked in black. Corresponding 95\% confidence intervals are shown in red.}
    \label{fig:s2_plots}
\end{figure}
\begin{figure}[h]
  \centering
  \includegraphics[width=0.7\textwidth]{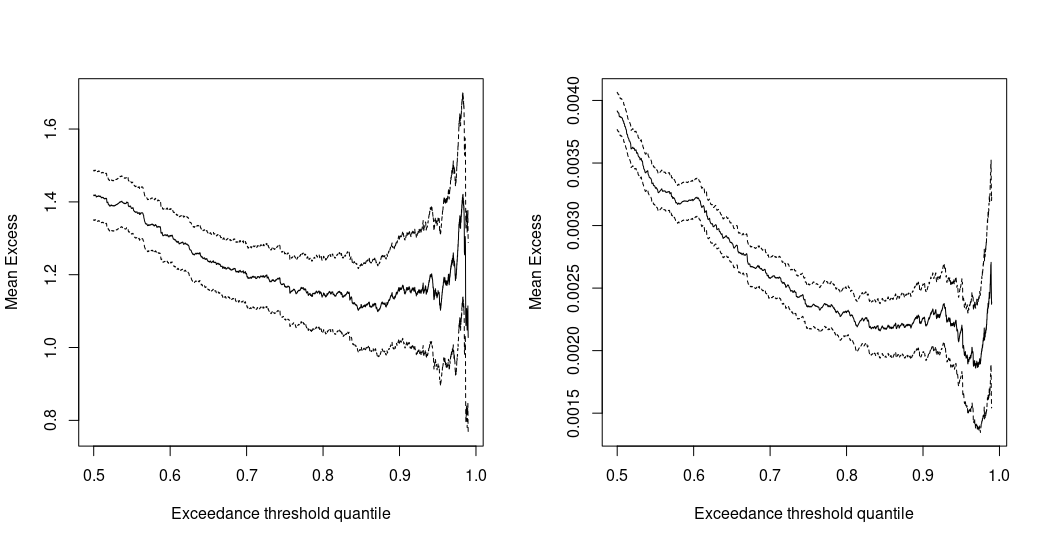}
    \caption{Mean residual life (mean excess) plots for excesses of $H_S$ and $S_2$. Point estimates and 95\% confidence intervals of mean excess with given exceedance probability are shown. The GPD implies a linear trend in mean excesses with respect to threshold. Thus, exceedance probabilities for which this linear trend is observed provide valid exceedance thresholds.}
    \label{fig:mrl_plots}
\end{figure}

\section{Wave field simulation}

\subsection{Airy wave theory} \label{section:lin_wave_extra}

Section~3.3.2 introduces the wave model of \cite{TylEA97} which is itself derived from the work of \cite{airy1845tides}, commonly referred to as linear wave theory, which provides physically based models for the surface elevation of the wave surface and associated kinematics. \cite{airy1845tides} models the stochastic surface elevation $E$ at time $t\in \RR$ and location $x \in \RR$ as 
\begin{equation} \label{equation:eta}
    E(t; x)=\sum_{n=1}^N\left\{A_n \cos \left(\omega_n t- k_n x\right)+B_n\sin \left(\omega_n t - k_n x\right)\right\},
\end{equation}
 with contributing angular frequencies $\omega_n>0$, coefficients $A_n,B_n>0$ and wave numbers $k_n$ as defined in Section~3.3.2, for $n=1,\ldots, N$. This is the non-conditioned form of expression~(10), for an arbitrary location $x \in \RR$. This model ensures conservation of mass, momentum satisfaction of necessary boundary conditions in a simple setting; see \cite{holthuijsen2010waves} for details.
 
 \cite{airy1845tides} also introduces the `velocity potential' equation
\begin{equation}\label{equation:velocity_potential}
    \phi(t;x, z)=\sum_{n=1}^N\left\{\frac{\omega_n \cosh [k_n(d+z)]}{k_n\sinh (k_n d)}\left[ -A_n\sin (\omega_n t- k_n x) + B_n\cos (\omega_n t- k_n x)\right]\right\},
\end{equation}
for time $t \in \RR$, horizontal position $x \in \RR$ and vertical position $z \in \RR$ relative to mean surface level, with water depth $d$. As in the main text, all physical quantities are given in SI units. Expression~\eqref{equation:velocity_potential} possesses the property 
\begin{equation} \label{equation:velocity_potential_diff}
    U(t; x, z)=\frac{\partial \phi}{\partial x} \quad \text{and} \quad V(t; x, z) = \frac{\partial \phi}{ \partial z},
\end{equation}
where $U(t; x, z)$ and $V(t;x, z)$ are horizontal and vertical velocities respectively. Hence, differentiation of \eqref{equation:velocity_potential} yields 
\begin{align}
    U(t;x, z) &= \sum_{n=1}^N \left[ \left\{A_n \cos(\omega_n t  - k_n x) + B_n \sin(\omega_n t  - k_n x)\right\} \omega_n \frac{\cosh\left[k_n(d+z)\right]}{\sinh(k_nd)}\right], \\
    V(t; x, z) &= \sum_{n=1}^N \left[ \left\{-A_n \sin(\omega_n t - k_n x) + B_n \cos(\omega_n t  - k_n x)\right\} \omega_n \frac{\sinh\left[k_n(d+z)\right]}{\sinh(k_nd)}\right], \\
    \dot{U}(t;x,  z) &= \sum_{n=1}^N \left[ \left\{-A_n \sin(\omega_n t - k_n x) + B_n \cos(\omega_n t  - k_n x)\right\} \omega_n^2 \frac{\cosh\left[k_n(d+z)\right]}{\sinh(k_nd)}\right], \\
    \dot{V}(t; x,  z) &= \sum_{n=1}^N \left[ \left\{-A_n \cos(\omega_n t - k_n x) - B_n \sin(\omega_n t - k_n x )\right\} \omega_n^2\frac{\cosh\left[k_n(d+z)\right]}{\sinh(k_nd)}\right], \label{equation:kinematics}\\
\end{align}
for horizontal velocity $U$, horizontal acceleration $\dot{U}$, vertical velocity $V$ and vertical acceleration $\dot{V}$. Note that in the main text we only utilise models derived from the above expressions for the horizontal velocity $U$ and acceleration $\dot{U}$ with $x=0$; the expressions for $V$ and $\dot{V}$ are included here for completeness.

\subsection{JONSWAP wave spectrum} \label{section:JONSWAP}

In Section~3.3.2, we assume the wave spectrum $S(\omega; \mathbf{X})$ has a JONSWAP parametric form, which is shown by \cite{hasselmann1973measurements} to be suitably flexible to capture measure offshore spectral behaviour. The JONSWAP spectrum has spectral density function 
$$
S(\omega ; \mathbf{X}=\mathbf{x})=\alpha \omega^{-r} \exp \left\{-\frac{r}{4}\left(\frac{|\omega|}{\omega_p(\mathbf{x})}\right)^{-4}\right\} \gamma \delta(\omega ; \mathbf{x}),
$$
for $\omega>0$, where $\mathbf{X}=(H_s, S_2)$ and $\omega_p(\mathbf{x}) = 2\pi/t(\mathbf{x})$, where $t(\mathbf{x})$ is the observed value of the second spectral moment wave period $T_2 = \sqrt{({2\pi H_S})/({g S_2})}$ in sea state $\mathbf{X}=\mathbf{x}$, with 
$$
\delta(\omega ; \mathbf{x})=\exp \left\{-\frac{1}{2\left(0.07+0.02 \cdot I\{{\omega_p(\mathbf{x})}>|\omega|\}\right)^2}\left(\frac{|\omega|}{\omega_p(\mathbf{x})}-1\right)^2\right\},
$$
and constants $r, \alpha, \gamma > 0$. The normalising constant $\alpha$ is chosen so that 
$$
4\cdot\left\{\int_{-\infty}^\infty S(\omega; \mathbf{x})\wrt{\omega}\right\}^{\frac{1}{2}} = h(\mathbf{x}),
$$
where $h(\mathbf{x})$ is the observed value of significant wave height $H_S$ in sea state $\mathbf{X}=\mathbf{x}$.

\subsection{Efficient wave simulation using the fast Fourier transform}

We apply the fast Fourier transform (FFT) algorithm \citep{nussbaumer1982fast} to efficiently compute the linear wave behaviour given by expressions (10) and (11) of Section~3.3.2, by formulating them as the discrete Fourier transforms of appropriately chosen `link functions'. Specifically, for 
$$
A'_n = A_n + \mathcal{Q}\sigma^2_n \quad \text{and} \quad B'_n = B_n + \mathcal{R}\sigma^2_n \omega_n,
$$
we write
\begin{align*}
    E(t_j) &= \sum^{N}_{n=1} g^{(\eta)}_n \exp\left\{{\frac{-2\pi i}{N}jn}\right\}, \quad g^{(\eta)}_n = A'_n + B'_n i, \\
    U(t_j, z) &= \sum^{N}_{n=1} g^{(u)}_n \exp\left\{{\frac{-2\pi i}{N}jn}\right\}, \quad  g^{(u)}_n = (A'_n + B'_n i) \cdot \omega_n  \cdot \frac{\cosh\{k_n(z+d)\}}{\sinh(k_nd)},\\
    \dot{U}(t_j, z) &= \sum^{N}_{n=1} g^{(\dot{u})}_n \exp\left\{{\frac{-2\pi i}{N}jn}\right\}, \quad g^{(\dot{u})}_n = (B'_n-A'_ni) \cdot \omega_n^2 \cdot \frac{\cosh\{k_n(z+d)\}}{\sinh(k_nd)},
\end{align*}
where $i^2 = -1$ and $j=1, \ldots, N$. This allows us to utilise the fast Fourier transform \citep{nussbaumer1982fast} to efficiently compute the linear wave behaviour given by expressions (10) and (11) for $z \in \RR$ at times $\{t_j\}_{j=1}^N$. Due to the intractability of the above kinematics at values of $z>0$, we employ simple linear kinematic stretching and evaluate the above equations at 
$$
z' = \begin{cases}
    z & \text{for}\,\, z\leq 0, \\
    0 & \text{for}\,\, z>0,
\end{cases}
$$
rather than $z$.

\section{Storm peak hierarchical model selection} \label{section:cv_scores_full}
\subsection{Distributional forms for steepness modelling}

Here we give the parametric forms of the four distributions considered for the modelling of $S_2|H_S$ and $S^-_2|H_S$ in Section~4.3, described by the most appropriate of their distribution and density functions. First, is the Generalised Extreme Value (GEV) distribution, which has distribution function
\begin{equation} 
F_\text{GEV}(x)=\exp \left\{-\left[1+\xi\left(\frac{x-\mu_G}{\sigma_G}\right)\right]^{-1 / \xi}\right\},
\end{equation}
defined on the set $\{x: 1+\xi(x-\mu) / \sigma>0\}$, with location $\mu_G \in \RR$, scale $\sigma_G >0$ and shape $\xi \in \RR$. Second, the Weibull distribution, which has distribution function
\begin{equation}
    F_\text{Weibull}(x)=1-e^{-(x / \lambda)^k},
\end{equation}
for $x \geq 0$, with scale $\lambda>0$ and shape $k>0$. Third, the Lognormal distribution, with density function
\begin{equation}
f_\text{Lognormal}(x) = \frac{1}{x \sigma_L \sqrt{2 \pi}} \exp \left(-\frac{(\ln x-\mu_L)^2}{2 \sigma_L^2}\right),
\end{equation}
for $x >0$, with location $\mu_L$ and log-scale $\sigma_L>0$. Fourth, the gamma distribution, with density function
\begin{equation}
    f_\text{Gamma}(x) = \frac{\beta^\alpha}{\Gamma(\alpha)} x^{\alpha-1} e^{-\beta x},
\end{equation}
where $\Gamma$ is the gamma function, for $x \geq 0$, with shape $\alpha>0$ and rate $\beta>0$.  

\subsection{Detailed cross validation results}
Here we present the full details of the results summarised by Table~1 in Section~5.2, summarised in Table~\ref{table:big_results} which shows the AIC, cross validation scores and AS for all 72 candidate models for $S_2|H_S$ and $S^-_2|H_S$. The distributional form and function of $H_S$ imposed on distribution parameters are given, as well as individual scores for each of the assessment criteria. Columns titled `$p_v$CV$K$' contain $K$-fold cross validation scores for an extreme threshold $v= \tilde{F}_{S_2} (p_v)$. The AS score is obtained by averaging of AIC and cross validation scores. Each cross validation scenario is repeated for 30 replicates for each data set, allowing the calculation of standard errors.  The highest ranking model per distribution per scoring method is marked in bold. In the event of a tie, we opt for the model with the highest AIC. 

\begin{table}[h]
\centering
\resizebox{\textwidth}{!}{\begin{tabular}{llllllllll}
  \hline
Distribution & Parameter Forms & AIC & 0.0CV5 & 0.0CV10 & 0.8CV5 & 0.8CV10 & 0.9CV10 & 0.9CV10 & Aggregate Score (AS)\\
  \hline
$S^-_2|H_S \sim \text{GEV}$ & Exp $\mu_G$, Exp $\sigma_G$ & \textbf{3.931} & 3.931 (0.002) & 3.931 (0.001) & 4.032 (0.014) & 4.037 (0.004) & 4.024 (0.002) & \textbf{4.11 (0.002)} & \textbf{3.999 (0.002)} \\ 
$S^-_2|H_S \sim \text{GEV}$& Exp $\mu_G$, Lin $\sigma_G$ & 3.930 & \textbf{3.937 (0.004)} & \textbf{3.936 (0.003)} & 3.945 (0.027) & 3.959 (0.014) & 3.998 (0.004) & 4.084 (0.005) & 3.97 (0.005) \\ 
 $S^-_2|H_S \sim \text{GEV}$ &Exp $\mu_G$, Qua $\sigma_G$ & 3.927 & 3.93 (0.001) & 3.93 (0.001) &\textbf{4.034 (0.007)} & 4.041 (0.003) & \textbf{4.025 (0.005)} & 4.106 (0.004) & \textbf{3.999 (0.002)} \\ 
 $S^-_2|H_S \sim \text{GEV}$ &  Lin $\mu_G$, Exp $\sigma_G$ & 3.918 & 3.919 (0.001) & 3.919 (0.001) & 4.012 (0.006) & 4.014 (0.005) & 3.975 (0.001) & 4.061 (0.001) & 3.974 (0.004) \\ 
  $S^-_2|H_S \sim \text{GEV}$& Lin $\mu_G$, Lin $\sigma_G$ & 3.913 & 3.926 (0.004) & 3.925 (0.003) & 3.953 (0.017) & 3.951 (0.01) & 3.975 (0.005) & 4.062 (0.001) & 3.958 (0.003) \\ 
  $S^-_2|H_S \sim \text{GEV}$ & Lin $\mu_G$, Qua $\sigma_G$ & 3.915 & 3.917 (0.001) & 3.916 (0.001) & 4.015 (0.007) & 4.016 (0.003) & 3.98 (0.004) & 4.06 (0.003) & 3.974 (0.001) \\ 
 $S^-_2|H_S \sim \text{GEV}$ &  Qua $\mu_G$, Exp $\sigma_G$ & 3.929 & 3.929 (0.004) & 3.93 (0.001) & 4.026 (0.012) & 4.029 (0.01) & 4.011 (0.013) & 4.101 (0.01) & 3.994 (0.003) \\ 
   $S^-_2|H_S \sim \text{GEV}$ & Qua $\mu_G$, Lin $\sigma_G$  & 3.927 & 3.934 (0.004) & 3.933 (0.004) & 3.955 (0.016) & 3.97 (0.016) & 3.992 (0.012) & 4.086 (0.008) & 3.971 (0.004) \\ 
  $S^-_2|H_S \sim \text{GEV}$ & Qua $\mu_G$, Qua $\sigma_G$ & 3.926 & 3.927 (0.003) & 3.927 (0.002) & 4.028 (0.012) & \textbf{4.043 (0.008)} & 4.002 (0.018) & 4.105 (0.007) & 3.994 (0.004) \\ 
   \hline
     $S^-_2|H_S \sim \text{Weibull}$ & Exp $k$, Exp $\lambda$ & \textbf{3.745} & \textbf{3.744 (0.001)} & \textbf{3.744 (0.001)} & \textbf{3.99 (0.003)} & \textbf{3.99 (0.002)} & 3.992 (0.008) & 4.075 (0.007) & \textbf{3.897 (0.002)} \\ 
     $S^-_2|H_S \sim \text{Weibull}$ & Exp $k$, Lin $\lambda$ & 3.679 & 3.666 (0.017) & 3.672 (0.006) & 3.647 (0.258) & 3.843 (0.035) & 3.656 (0.238) & 3.809 (0.088) & 3.71 (0.047) \\ 
  $S^-_2|H_S \sim \text{Weibull}$ & Exp $k$, Qua $\lambda$ & 3.711 & 3.726 (0.005) & 3.721 (0.003) & 3.955 (0.01) & 3.956 (0.009) & 3.942 (0.013) & 4.019 (0.015) & 3.861 (0.004) \\ 
   $S^-_2|H_S \sim \text{Weibull}$ & Lin $k$, Exp $\lambda$ & 3.743 & 3.743 (0.002) & 3.743 (0.001) & 3.982 (0.009) & 3.987 (0.004) & \textbf{3.993 (0.008)} & \textbf{4.081 (0.006)} & 3.896 (0.002) \\
    $S^-_2|H_S \sim \text{Weibull}$ & Lin $k$, Lin $\lambda$ & 3.578 & 3.584 (0.017) & 3.576 (0.02) & 3.828 (0.017) & 3.855 (0.008) & 3.829 (0.014) & 3.852 (0.015) & 3.729 (0.004) \\ 
  $S^-_2|H_S \sim \text{Weibull}$& Lin $k$, Qua $\lambda$& 3.697 & 3.695 (0.002) & 3.695 (0.002) & 3.885 (0.006) & 3.891 (0.007) & 3.807 (0.014) & 3.897 (0.015) & 3.795 (0.003) \\ 
   $S^-_2|H_S \sim \text{Weibull}$ & Qua $k$, Exp $\lambda$ & 3.732 & 3.734 (0.007) & 3.735 (0.005) & 3.944 (0.016) & 3.975 (0.009) & 3.95 (0.021) & 4.061 (0.012) & 3.876 (0.004) \\ 
   $S^-_2|H_S \sim \text{Weibull}$ & Qua $k$, Lin $\lambda$ & 3.718 & 3.715 (0.003) & 3.716 (0.002) & 3.905 (0.011) & 3.909 (0.009) & 3.859 (0.021) & 3.948 (0.021) & 3.824 (0.005) \\
   $S^-_2|H_S \sim \text{Weibull}$ & Qua $k$, Qua $\lambda$ & 3.712 & 3.697 (0.014) & 3.691 (0.014) & 3.87 (0.034) & 3.852 (0.028) & 3.832 (0.067) & 3.944 (0.027) & 3.8 (0.013) \\ 
     \hline
 $S^-_2|H_S \sim \text{Lognormal}$ & Exp $\mu_L$, Exp $\sigma_L$ & 3.887 & 3.855 (0.04) & 3.864 (0.026) & 3.765 (0.118) & 3.924 (0.054) & 3.664 (0.177) & 3.987 (0.108) & 3.849 (0.037) \\ 
  $S^-_2|H_S \sim \text{Lognormal}$ & Exp $\mu_L$, Lin $\sigma_L$ & 3.889 & 3.842 (0.04) & 3.802 (0.049) & 3.94 (0.039) & 3.928 (0.042) & 3.93 (0.076) & 3.909 (0.066) & 3.891 (0.018) \\ 
  $S^-_2|H_S \sim \text{Lognormal}$ & Exp $\mu_L$, Qua $\sigma_L$ & 3.891 & 3.886 (0.021) & 3.877 (0.029) & 3.976 (0.006) & 3.979 (0.011) & 3.997 (0.003) & 4.082 (0.004) & 3.955 (0.007) \\ 
   $S^-_2|H_S \sim \text{Lognormal}$ & Lin $\mu_L$, Exp $\sigma_L$ & 3.894 & 3.893 (0.002) & 3.893 (0.002) & 3.958 (0.008) & 3.964 (0.005) & 3.967 (0.01) & 4.052 (0.006) & 3.946 (0.002) \\ 
  $S^-_2|H_S \sim \text{Lognormal}$ & Lin $\mu_L$, Lin $\sigma_L$ & \textbf{3.896 }& \textbf{3.897 (0.001)} &\textbf{3.897 (0.001)} & \textbf{3.981 (0.004)} & \textbf{3.983 (0.003)} & \textbf{3.999 (0.003)} & \textbf{4.085 (0.002)} & \textbf{3.963 (0.001)} \\ 
   $S^-_2|H_S \sim \text{Lognormal}$ & Lin $\mu_L$, Qua $\sigma_L$ & 3.857 & 3.873 (0.01) & 3.868 (0.008) & 3.932 (0.026) & 3.911 (0.015) & 3.874 (0.038) & 3.935 (0.032) & 3.893 (0.010) \\
    $S^-_2|H_S \sim \text{Lognormal}$ & Qua $\mu_L$, Exp $\sigma_L$& 3.880 & 3.862 (0.037) & 3.874 (0.017) & 3.884 (0.063) & 3.843 (0.054) & 3.836 (0.057) & 3.875 (0.075) & 3.865 (0.019) \\ 
  $S^-_2|H_S \sim \text{Lognormal}$ & Qua $\mu_L$, Lin $\sigma_L$ & 3.891 & 3.892 (0.003) & 3.892 (0.001) & 3.97 (0.004) & 3.972 (0.003) & 3.976 (0.004) & 4.063 (0.002) & 3.951 (0.001) \\ 
 $S^-_2|H_S \sim \text{Lognormal}$ & Qua $\mu_L$, Qua $\sigma_L$& 3.881 & 3.879 (0.008) & 3.877 (0.006) & 3.925 (0.032) & 3.919 (0.035) & 3.95 (0.037) & 4.03 (0.019) & 3.923 (0.010) \\ 
   \hline
   $S^-_2|H_S \sim \text{Gamma}$ & Exp $\alpha$, Exp $\beta$ & 3.801 & 3.858 (0.009) & 3.854 (0.009) & 4.016 (0.023) & 3.998 (0.014) & 3.917 (0.035) & 4.079 (0.011) & 3.932 (0.007) \\ 
  $S^-_2|H_S \sim \text{Gamma}$& Exp $\alpha$, Lin $\beta$ & 3.775 & 3.815 (0.021) & 3.821 (0.016) & 3.928 (0.021) & 3.923 (0.015) & 3.832 (0.027) & 3.921 (0.037) & 3.859 (0.007) \\ 
  $S^-_2|H_S \sim \text{Gamma}$ & Exp $\alpha$, Qua $\beta$ & \textbf{3.876} & 3.864 (0.006) & 3.863 (0.004) & 3.971 (0.02) & 3.965 (0.015) & 3.915 (0.023) & 3.982 (0.02) & 3.919 (0.005) \\
    $S^-_2|H_S \sim \text{Gamma}$ & Lin $\alpha$, Exp $\beta$& 3.872 & 3.872 (0.002) & 3.873 (0.001) & \textbf{4.02 (0.005)} & \textbf{4.021 (0.005)} & 3.991 (0.007) & 4.08 (0.006) & 3.961 (0.002) \\ 
$S^-_2|H_S \sim \text{Gamma}$ & Lin $\alpha$, Lin $\beta$ & 3.871 & \textbf{3.872 (0.002)} & \textbf{3.873 (0.001)} & 4.008 (0.005) & 4.01 (0.008) & 3.974 (0.006) & 4.06 (0.005) & 3.953 (0.002) \\ 
$S^-_2|H_S \sim \text{Gamma}$ & Lin $\alpha$, Qua $\beta$ & 3.866 & 3.868 (0.001) & 3.868 (0) & 4.001 (0.005) & 4.004 (0.003) & 3.959 (0.004) & 4.046 (0.004) & 3.945 (0.001) \\ 
   $S^-_2|H_S \sim \text{Gamma}$ & Qua $\alpha$, Exp $\beta$ & 3.823 & 3.826 (0.004) & 3.825 (0.003) & 3.94 (0.052) & 3.947 (0.01) & 3.881 (0.022) & 3.955 (0.016) & 3.885 (0.009) \\ 
 $S^-_2|H_S \sim \text{Gamma}$ & Qua $\alpha$, Lin $\beta$ & 3.875 & 3.866 (0.007) & 3.866 (0.006) & 4.017 (0.019) & 4.019 (0.016) & \textbf{4.015 (0.017)} & \textbf{4.081 (0.025)} & \textbf{3.963 (0.006)} \\ 
 $S^-_2|H_S \sim \text{Gamma}$ & Qua $\alpha$, Qua $\beta$ & 3.837 & 3.844 (0.007) & 3.845 (0.006) & 3.876 (0.024) & 3.918 (0.026) & 3.823 (0.051) & 3.9 (0.046) & 3.863 (0.012) \\ 
   \hline
    $S_2|H_S \sim \text{GEV}$ & Exp $\mu_G$, Exp $\sigma_G$ & 3.567 & 3.446 (0.035) & 3.479 (0.021) & \textbf{3.822 (0.044)} & \textbf{3.862 (0.023)} & \textbf{3.919 (0.01)} & \textbf{3.973 (0.009)} & 3.724 (0.010) \\ 
   $S_2|H_S \sim \text{GEV}$& Exp $\mu_G$, Lin $\sigma_G$ & 3.495 & 3.399 (0.023) & 3.42 (0.013) & 3.586 (0.026) & 3.561 (0.031) & 3.677 (0.013) & 3.734 (0.009)  & 3.553 (0.007) \\ 
   $S_2|H_S \sim \text{GEV}$& Exp $\mu_G$, Qua $\sigma_G$ & 3.555 & 3.532 (0) & 3.531 (0) & 3.82 (0.001) & 3.816 (0.001) & 3.875 (0.001) & 3.954 (0) &\textbf{3.726 (0.000)} \\ 
   $S_2|H_S \sim \text{GEV}$& Lin $\mu_G$, Exp $\sigma_G$ & 3.370 & 3.305 (0.016) & 3.316 (0.012) & 3.537 (0.105) & 3.25 (0.075) & 3.025 (0.178) & 2.674 (0.088) & 3.211 (0.032) \\ 
   $S_2|H_S \sim \text{GEV}$ & Lin $\mu_G$, Lin $\sigma_G$ & 3.655 & 3.551 (0.009) & 3.555 (0.013) & 3.645 (0.052) & 3.619 (0.041) & 3.29 (0.172) & 3.482 (0.032) & 3.542 (0.028) \\ 
   $S_2|H_S \sim \text{GEV}$ & Lin $\mu_G$, Qua $\sigma_G$ & \textbf{3.688} & \textbf{3.697 (0.002)} & \textbf{3.694 (0.001)} & 3.558 (0.003) & 3.554 (0.002) & 3.237 (0.004) & 3.305 (0.002) & 3.533 (0.001) \\ 
   $S_2|H_S \sim \text{GEV}$ & Qua $\mu_G$, Exp $\sigma_G$ & 3.638 & 3.436 (0.067) & 3.567 (0.008) & 3.67 (0.019) & 3.673 (0.017) & 3.575 (0.032) & 3.586 (0.024) & 3.592 (0.012) \\
  $S_2|H_S \sim \text{GEV}$ & Qua $\mu_G$, Lin $\sigma_G$  & 3.287 & 3.104 (0.213) & 3.256 (0.089) & 1.966 (0.087) & 1.686 (0.037) & 1.344 (0.017) & 1.323 (0) & 2.281 (0.038) \\
     $S_2|H_S \sim \text{GEV}$ & Qua $\mu_G$, Qua $\sigma_G$ & 3.356 & 2.89 (0.129) & 3.094 (0.06) & 2.365 (0.205) & 2.535 (0.122) & 1.979 (0.149) & 2.567 (0.06) & 2.684 (0.047) \\ 
   \hline
   $S_2|H_S \sim \text{Weibull}$ & Exp $k$, Exp $\lambda$ & 3.864 & 3.867 (0.002) & 3.868 (0.002) & 3.966 (0.007) & 3.979 (0.005) & 3.87 (0.011) & 3.978 (0.013) & 3.913 (0.003) \\ 
  $S_2|H_S \sim \text{Weibull}$ & Exp $k$, Lin $\lambda$ & 3.880 & 3.885 (0.002) & 3.885 (0.002) & 4.022 (0.006) & 4.024 (0.005) & 3.961 (0.009) & 4.05 (0.007) & 3.958 (0.002) \\ 
  $S_2|H_S \sim \text{Weibull}$ & Exp $k$, Qua $\lambda$ & 3.897 & 3.9 (0.002) & 3.9 (0.001) & 4.04 (0.006) & 4.041 (0.004) & 3.995 (0.02) & 4.088 (0.005) & 3.98 (0.003) \\ 
    $S_2|H_S \sim \text{Weibull}$ & Lin $k$, Exp $\lambda$ & 3.872 & 3.873 (0.001) & 3.874 (0.001) & 3.984 (0.005) & 3.989 (0.004) & 3.902 (0.004) & 3.994 (0.006) & 3.927 (0.002) \\ 
   $S_2|H_S \sim \text{Weibull}$ & Lin $k$, Lin $\lambda$ & 3.888 & 3.891 (0.003) & 3.891 (0.002) & 4.008 (0.012) & 4.013 (0.009) & 3.96 (0.007) & 4.022 (0.015) & 3.953 (0.003) \\ 
 $S_2|H_S \sim \text{Weibull}$ & Lin $k$, Qua $\lambda$ & 3.901 & \textbf{3.902 (0.006)} & \textbf{3.903 (0.003)} & 4.032 (0.006) & 4.038 (0.006) &\textbf{4.006 (0.007)} & \textbf{4.096 (0.008)} & \textbf{3.983 (0.003)} \\ 
   $S_2|H_S \sim \text{Weibull}$ & Qua $k$, Exp $\lambda$ & 3.871 & 3.872 (0.001) & 3.871 (0.001) & 3.978 (0.009) & 3.986 (0.006) & 3.892 (0.009) & 3.989 (0.006) & 3.923 (0.003) \\ 
 $S_2|H_S \sim \text{Weibull}$ & Qua $k$, Lin $\lambda$ & 3.889 & 3.89 (0.002) & 3.89 (0.001) & 4.018 (0.005) & 4.021 (0.002) & 3.957 (0.005) & 4.047 (0.004) & 3.959 (0.001) \\ 
$S_2|H_S \sim \text{Weibull}$ & Qua $k$, Qua $\lambda$ & \textbf{3.903} & 3.88 (0.044) & 3.9 (0.01) & \textbf{4.046 (0.003)} & \textbf{4.043 (0.002)} & \textbf{4.006 (0.003)} & \textbf{4.096 (0.005) }& 3.982 (0.007) \\ 
   \hline
   $S_2|H_S \sim \text{Lognormal}$ & Exp $\mu_L$, Exp $\sigma_L$ & 3.629 & 3.632 (0.007) & 3.636 (0.005) & 3.872 (0.021) & 3.869 (0.01) & 3.8 (0.022) & 3.891 (0.025) & 3.761 (0.006) \\ 
    $S_2|H_S \sim \text{Lognormal}$& Exp $\mu_L$, Lin $\sigma_L$ & 3.485 & 3.485 (0.006) & 3.497 (0.006) & 3.803 (0.009) & 3.807 (0) & 3.861 (0) & 3.941 (0) & 3.697 (0.002) \\ 
  $S_2|H_S \sim \text{Lognormal}$& Exp $\mu_L$, Qua $\sigma_L$ & 3.563 & 3.562 (0.002) & 3.562 (0.001) & 3.896 (0.004) & 3.897 (0.002) & 3.926 (0.006) & \textbf{4.008 (0.004)} & 3.773 (0.001) \\ 
 $S_2|H_S \sim \text{Lognormal}$ & Lin $\mu_L$, Exp $\sigma_L$ & 3.651 & 3.649 (0.003) & 3.65 (0.002) & 3.914 (0.008) & 3.92 (0.007) & 3.878 (0.011) & 3.961 (0.016) & 3.803 (0.003) \\
  $S_2|H_S \sim \text{Lognormal}$ & Lin $\mu_L$, Lin $\sigma_L$ & 3.591 & 3.59 (0.003) & 3.588 (0.003) & 3.69 (0.016) & 3.705 (0.012) & 3.584 (0.027) & 3.7 (0.015) & 3.635 (0.005) \\ 
  $S_2|H_S \sim \text{Lognormal}$ & Lin $\mu_L$, Qua $\sigma_L$ & 3.612 & 3.613 (0.001) & 3.613 (0.001) & 3.747 (0.004) & 3.754 (0.003) & 3.563 (0.004) & 3.646 (0.006) & 3.65 (0.001) \\ 
 $S_2|H_S \sim \text{Lognormal}$ & Qua $\mu_L$, Exp $\sigma_L$ & \textbf{3.669} & \textbf{3.657 (0.007)} & \textbf{3.661 (0.004)} & \textbf{3.942 (0.01)} & 3.925 (0.016) & \textbf{3.947 (0.013)} & 3.97 (0.028) & \textbf{3.824 (0.005)} \\ 
  $S_2|H_S \sim \text{Lognormal}$ & Qua $\mu_L$, Lin $\sigma_L$ & 3.659 & 3.653 (0.011) & 3.653 (0.01) & 3.895 (0.062) & \textbf{3.928 (0.01)} & 3.895 (0.055) & 3.901 (0.326) & 3.798 (0.046) \\ 
$S_2|H_S \sim \text{Lognormal}$ & Qua $\mu_L$, Qua $\sigma_L$ & 3.185 & 3.227 (0.021) & 3.212 (0.016) & 2.642 (0.074) & 2.626 (0.035) & 2.191 (0.099) & 2.37 (0.061) & 2.779 (0.765) \\ 
   \hline
   $S_2|H_S \sim \text{Gamma}$ & Exp $\alpha$, Exp $\beta$ & 3.608 & 3.606 (0.008) & 3.605 (0.005) & 3.694 (0.011) & 3.685 (0.01) & 3.66 (0.017) & 3.773 (0.011) & 3.662 (0.003) \\ 
  $S_2|H_S \sim \text{Gamma}$ & Exp $\alpha$, Lin $\beta$ & 3.669 & 3.643 (0.006) & 3.645 (0.005) & 3.795 (0.01) & 3.802 (0.006) & 3.752 (0.013) & 3.843 (0.016) & 3.736 (0.004) \\ 
  $S_2|H_S \sim \text{Gamma}$ & Exp $\alpha$, Qua $\beta$ & 3.604 & 3.602 (0.015) & 3.601 (0.013) & 3.974 (0.014) & 3.974 (0.015) & 3.96 (0.016) & 4.038 (0.028) & 3.822 (0.004) \\ 
   $S_2|H_S \sim \text{Gamma}$ & Lin $\alpha$, Lin $\beta$ & \textbf{3.710} & \textbf{3.709 (0.004)} & \textbf{3.71 (0.002)} & 3.975 (0.002) & 3.975 (0.003) & 3.972 (0.007) & \textbf{4.058 (0.006)} & \textbf{3.873 (0.002)} \\ 
     $S_2|H_S \sim \text{Gamma}$ & Lin $\alpha$, Exp $\beta$ & 3.668 & 3.668 (0.003) & 3.668 (0.003) & 3.85 (0.008) & 3.857 (0.005) & 3.816 (0.007) & 3.902 (0.007) & 3.776 (0.002) \\ 
 $S_2|H_S \sim \text{Gamma}$  & Lin $\alpha$, Qua $\beta$ & 3.692 & 3.665 (0.025) & 3.662 (0.024) & 3.91 (0.024) & 3.911 (0.019) & 3.898 (0.045) & 3.995 (0.018) & 3.819 (0.009) \\ 
   $S_2|H_S \sim \text{Gamma}$ & Qua $\alpha$, Exp $\beta$ & 3.599 & 3.487 (0.138) & 3.433 (0.112) & 3.904 (0.071) & 3.938 (0.011) & 3.922 (0.01) & 3.987 (0.026) & 3.753 (0.027) \\ 
 $S_2|H_S \sim \text{Gamma}$  & Qua $\alpha$, Lin $\beta$ & 3.671 & 3.67 (0.007) & 3.668 
 (0.005) & 3.908 (0.008) & 3.898 (0.005) & 3.836 (0.015) & 3.993 (0.011) & 3.806 (0.003) \\ 
   $S_2|H_S \sim \text{Gamma}$ & Qua $\alpha$, Qua $\beta$ & 3.622 & 3.618 (0.005) & 3.618 (0.004) & \textbf{3.985 (0.004)} & \textbf{3.985 (0.003)} & \textbf{3.974 (0.005)} & 4.045 (0.014) & 3.835 (0.003) \\ 
   \hline
\end{tabular}}
\caption{AIC and cross validation scores of all model combinations for $S_2|H_S$ and $S^-_2|H_S$, for all considered distribution parameter functional forms. The highest scoring model per distribution per column is marked in bold. Sample standard errors are given to three decimal places.}
\label{table:big_results}
\end{table}

\end{document}